%% file: main.tex
\documentclass[%
reprint,
twocolumn,
 amsmath,amssymb,
aps,
prl,
]{revtex4-2}

\usepackage{graphicx}% Include figure files
\usepackage{physics}
\usepackage{siunitx}
\usepackage{algpseudocode}
\usepackage{libertine}
\usepackage{longtable}
\usepackage{booktabs}

\newcommand{\tstar}{T_{2}^{\star}}
\newcommand{\ttwo}{T_{2}}

\newcommand{\Bz}{B_{\mathrm{z}}}
\newcommand{\Azz}{A_{\mathrm{zz}}}
\newcommand{\Azx}{A_{\mathrm{zx}}}

\newcommand{\Iz}[1]{\hat{I}^{(#1)}_\mathrm{z}}
\newcommand{\Ix}[1]{\hat{I}^{(#1)}_\mathrm{x}}
\newcommand{\Iy}[1]{\hat{I}^{(#1)}_\mathrm{y}}
\newcommand{\Ia}[1]{\hat{I}^{(#1)}_\mathrm{\alpha}}

\newcommand{\sa}[1]{\hat{\sigma}^{(#1)}_\mathrm{\alpha}}
\newcommand{\phip}[1]{\phi_{\perp}^{(#1)}}
\newcommand{\ide}{\hat{\mathbb{I}}}
\newcommand{\pitwox}[1]{{R_{\mathrm{x}}(\tfrac{\pi}{2})}^{(#1)}}

\newcommand{\pitwok}[1]{{R_{\mathrm{k}}(\tfrac{\pi}{2})}^{(#1)}}

\newcommand{\pix}[1]{{R_{\mathrm{x}}(\pi)}^{(#1)}}
\newcommand{\Rk}[1]{{R_{\mathrm{k}}(\theta)}^{(#1)}}
\newcommand{\Rx}[2]{{R_{\mathrm{x}}(#2)}^{(#1)}}
\newcommand{\gc}{\gamma_\mathrm{c}}
\newcommand{\gee}{\gamma_\mathrm{e}}
\newcommand{\Uint}[1]{U_{\mathrm{int}}}
\newcommand{\Up}{U_{+}}
\newcommand{\Um}{U_{-}}
\newcommand{\Uxx}[2]{U_{\mathrm{xx}}^{(#1\rightarrow #2)}}

\newcommand{\Uxk}[2]{U_{\mathrm{kx}}^{(#1\rightarrow #2)}}

\newcommand{\aj}[2]{\alpha^{#1#2}_j}
\newcommand{\bj}[2]{\beta^{#1#2}_j}
\newcommand{\bjs}[2]{ {\beta^{#1#2}_j}^* }
\newcommand{\deltah}[1]{\Delta_{#1}}

\begin{document}

\title{Mapping a 50-spin-qubit network through correlated sensing}

\author{G.L. van de Stolpe$^{1,2}$}
\author{D. P. Kwiatkowski$^{1,2}$}
\author{C.E. Bradley$^{1,2}$}
\author{J. Randall$^{1,2}$}
\author{M.H. Abobeih$^{1,2}$}
\author{S. A. Breitweiser$^{3}$}
\author{L. C. Bassett$^{3}$}
\author{M. Markham$^{4}$}
\author{D.J. Twitchen$^{4}$}
\author{T.H. Taminiau$^{1,2}$}
% \email{t.h.taminiau@tudelft.nl}

\affiliation{$^1$QuTech, Delft University of Technology, PO Box 5046, 2600 GA Delft, The Netherlands}%

\affiliation{$^2$Kavli Institute of Nanoscience Delft, Delft University of Technology,
PO Box 5046, 2600 GA Delft, The Netherlands}

\affiliation{$^3$Quantum Engineering Laboratory, Department of Electrical and Systems Engineering, University of Pennsylvania, 200 South 33rd Street, Philadelphia, PA 19104, USA}

\affiliation{$^4$Element Six Innovation, Fermi Avenue, Harwell Oxford, Didcot, Oxfordshire OX11 0QR, United Kingdom}

\begin{abstract}
    Spins associated to optically accessible solid-state defects have emerged as a versatile platform for exploring quantum simulation, quantum sensing and quantum communication. Pioneering experiments have shown the sensing, imaging, and control of multiple nuclear spins surrounding a single electron-spin defect. However, the accessible size of these spin networks has been constrained by the spectral resolution of current methods. Here, we map a network of 50 coupled spins through high-resolution correlated sensing schemes, using a single nitrogen-vacancy center in diamond. We develop concatenated double-resonance sequences that identify spin-chains through the network. These chains reveal the characteristic spin frequencies and their interconnections with high spectral resolution, and can be fused together to map out the network. Our results provide new opportunities for quantum simulations by increasing the number of available spin qubits. Additionally, our methods might find applications in nano-scale imaging of complex spin systems external to the host crystal.
\end{abstract}

\maketitle

\section{Introduction}
Optically interfaced spin qubits associated to defects in solids provide a versatile platform for quantum simulation \cite{randall_manybody_2021}, quantum networks \cite{pompili_realization_2021,hermans_qubit_2022} and quantum sensing \cite{degen_quantum_2017,casola_probing_2018, janitz_diamond_2022}. Various systems are being explored \cite{awschalom_quantum_2018}, including defects in diamond \cite{randall_manybody_2021, pompili_realization_2021,hermans_qubit_2022, debroux_quantum_2021, sipahigil_integrated_2016}, silicon carbide \cite{bourassa_entanglement_2020,lukin_integrated_2020}, silicon \cite{durand_broad_2021, higginbottom_optical_2022}, hexagonal boron nitride (hBN) \cite{gao_nuclear_2022}, and rare-earth ions \cite{ruskuc_nuclear_2022}. The defect electron spin provides a qubit with high-fidelity control, optical initialization and readout, and a (long-range) photonic quantum network interface \cite{pompili_realization_2021, hermans_qubit_2022}. Additionally, the electron spin can be used to sense and control multiple nuclear spins surrounding the defect \cite{bradley_tenqubit_2019,babin_fabrication_2022, ruskuc_nuclear_2022}. This additional network of coupled spins provides a qubit register for quantum information processing, as well as a test bed for nanoscale magnetic resonance imaging \cite{abobeih_atomicscale_2019, cujia_parallel_2022, zhao_atomicscale_2011, schwartz_blueprint_2019, perunicic_quantum_2016, wang_positioning_2016}. Examples of emerging applications are quantum simulations of many-body physics \cite{unden_coherent_2018, cai_largescale_2013, randall_manybody_2021,geng_ancilla_2021, davis_probing_2023}, as well as quantum networks \cite{pompili_realization_2021,hermans_qubit_2022}, where the nuclear spins provide qubits for quantum memory \cite{bartling_entanglement_2022}, entanglement distillation \cite{kalb_entanglement_2017}, and error correction \cite{waldherr_quantum_2014, taminiau_universal_2014,abobeih_faulttolerant_2022}.

State-of-the-art experiments have demonstrated the imaging of spin networks containing up to 27 nuclear spins \cite{abobeih_atomicscale_2019,cujia_parallel_2022,shi_singleprotein_2015,zopes_reconstructionfree_2019,zopes_threedimensional_2018}. The ability to map larger spin networks can be a precursor for quantum simulations that are currently intractable, would provide a precise understanding of the noise environment of spin-qubit registers \cite{abobeih_faulttolerant_2022, xie_99_2023, wise_using_2021}, and might contribute towards efforts to image complex spin systems outside of the host material \cite{perunicic_quantum_2016, lovchinsky_nuclear_2016, wang_positioning_2016,zhao_atomicscale_2011, schwartz_blueprint_2019, ajoy_atomicscale_2015}. A key open challenge for mapping larger networks is spectral crowding, which causes overlapping signals and introduces ambiguity in the assignment of signals to individual spins and the interactions between them.

Here, we develop correlated sensing sequences that measure both the network connectivity as well as the characteristic spin frequencies with high spectral resolution. We apply these sequences to map a 50-nuclear-spin network comprised of 1225 spin-spin interactions in the vicinity of a nitrogen vacancy (NV) center in diamond. The key concept of our method is to concatenate double-resonance sequences to measure chains of coupled spins through the network. The mapping of spin chains removes ambiguity about how the spins are connected and enables the sensing of spins that are farther away from the electron-spin sensor in spectrally crowded regions. These results significantly increase the size and complexity of the accessible spin network. Additionally, our methods are applicable to a wide variety of systems, and might inspire future methods to magnetically image complex samples such as individual molecules or proteins \cite{perunicic_quantum_2016, shi_singleprotein_2015, ajoy_atomicscale_2015}.

\section{Results}
\subsection{Spin-network mapping}\label{sec:spin_network_mapping}
\begin{figure*}
    \centering
    \includegraphics[width=1\textwidth]{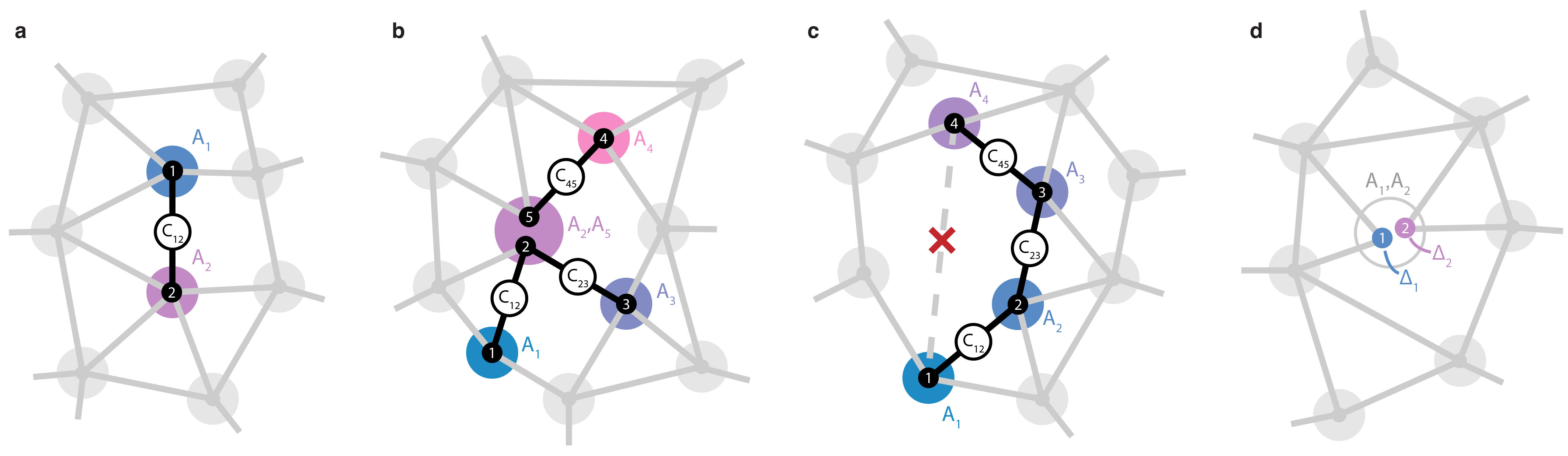}
    \caption{Mapping spin networks. Graph representing a spin network, where vertices denote spins and edges denote spin-spin interactions ($C_{ij}$). Spins are distributed among spectral regions (coloured disks) by their precession frequency ($A_{i}$). a) If all spin frequencies are unique (one spin in each disk), the network can be mapped by measuring only pairwise interactions ($C_{12}$) between frequencies ($A_1,A_2$). b) If spins spectrally overlap (e.g. spins 2 and 5 with $A_2 \approx A_5$) due to the finite line width set by the dephasing time $\tstar$, pairwise measurements alone are ambiguous when assigning interactions to specific spins. By measuring chains (e.g. through $A_1, A_2, A_3$) we directly retrieve the connectivity of the network. c) We also exploit spin chains to measure interactions between spins that are otherwise challenging to access. As an example, couplings belonging to Spin 4 are not directly accessible from the spin at $A_1$ – due to spectral crowding or negligible couplings – but can be obtained through a chain. d) Finally, we complement the spin chains with a correlated double-resonance method that enhances the spectral resolution for the spin-frequency shifts ($\Delta_i$) from $\sim 1/\tstar$ to $\sim 1/\ttwo$, so that spectrally overlapping spins can also be resolved directly. This figure shows a conceptual network with vertices organized in frequency space. In Supplementary Fig. 1, we discuss the specific relations between frequency and spatial position for the experimental system considered here: an NV center in diamond and surrounding $^{13}$C-spin network, for which increased spectral crowing (panels b-d) naturally occurs for $^{13}$C spins that are farther away from the NV center.  
    \label{fig:spin_networks}
    }
\end{figure*}

We consider a network of $N$ coupled nuclear spins in the vicinity of a single electron spin that acts as a quantum sensor \cite{cujia_parallel_2022,abobeih_atomicscale_2019}. The effective dynamics of the nuclear-spin network, with an external magnetic field along the z-axis, are described by the Hamiltonian (see Supplementary Note 1): 
\begin{equation}
\hat{H} = \sum^{N}_{i=1} A_i \Iz{i} 
+ \sum^{N}_{i=1} \sum^{N}_{j = i+1} C_{ij} \Iz{i} \,\Iz{j} \,,
\label{eq:nuclear_hamiltonian}
\end{equation}
where $\Iz{i}$ denotes the nuclear Pauli spin-$\tfrac{1}{2}$ operator for spin $i$, $A_i$ are the precession frequencies associated with each spin, and $C_{ij}$ denotes the nuclear-nuclear coupling between spin $i$ and $j$. The frequencies $A_i$ might differ due to differences in species (gyromagnetic ratio), the local magnetic field and spin environment, and due to coupling to the sensor electron spin. Our goal is to extract the characteristic spin frequencies $A_i$ and spin-spin couplings $C_{ij}$ that capture the structure of the network. 

Fig. \ref{fig:spin_networks} shows an example network, with coloured disks denoting frequency regions, and numbered dots inside signifying spins at these frequencies. Although in principle all spins are coupled to all spins, we draw edges only for strong, resolvable, spin-spin couplings, defined by: $C_{ij} \gtrsim 1/\ttwo$, where $\ttwo$ is the nuclear Hahn-echo coherence time ($\sim \SI{0.5}{\second}$) \cite{bradley_tenqubit_2019}. The network connectivity constitutes the presence (or absence) of such resolvable couplings. In general, the number of frequency disks is smaller than the number of spins, as multiple spins might occupy the same frequency region (i.e. overlap in frequency).

State-of-the-art spin-network mapping relies on isolating individual nuclear-nuclear interactions through spin-echo double resonance (SEDOR) \cite{abobeih_atomicscale_2019}. Applying simultaneous echo pulses at frequencies $A_i$ and $A_j$ preserves the interaction $C_{ij}$ between spins at $A_i$ and $A_j$, while decoupling them from (quasi-static) environmental noise and the rest of the network, so that the coupling $C_{ij}$ is encoded in the nuclear-spin polarisation with high spectral resolution (set by the nuclear $\ttwo$-time rather than $\tstar$-time). The signal is acquired by mapping the resulting nuclear spin polarisation, for example at frequency $A_i$, on the NV electron spin and reading it out optically \cite{bradley_tenqubit_2019}. Such a measurement yields a correlated list of three frequencies $\{A_i, C_{ij}, A_j\}$ (Fig. \ref{fig:spin_networks}a). If all spins are spectrally isolated, so that the $A_i$ do not overlap, these pairwise measurements completely characterise the network. 

However, due to their finite spectral line widths (set by $1/\tstar$), multiple spin frequencies $A_{i}$ may overlap (indicated by multiple spins occupying a disk). This introduces ambiguity when assigning measured couplings to specific spins in the network, and causes complex overlapping signals, which are difficult to resolve and interpret \cite{abobeih_atomicscale_2019,cujia_parallel_2022}. Figure \ref{fig:spin_networks}b shows an example where pairwise measurements break down; spins 2 and 5 overlap in frequency ($A_2 \approx A_5$). Applying pairwise SEDOR between frequencies $A_1, A_3, A_4$ and a frequency that overlaps with $A_2$ and $A_5$ returns three independent pairwise correlations: $\{A_1, C_{12}, A_2 \}, \{A_3, C_{23}, A_2 \} $ and $\{A_4, C_{45}, A_5\}$. Crucially, however, such measurements cannot distinguish this uncoupled 2-spin and 3-spin chain (Fig 1b) from a single 4-spin network (with a single central spin at $A_2$), nor from a network of 3 uncoupled 2-spin chains (three spectrally overlapping spins). Without introducing additional a-priori knowledge or assumptions about the system, pairwise measurements cannot be assigned to specific spins and are thus insufficient to reconstruct the network \cite{abobeih_atomicscale_2019}.   

Our approach is to measure connected chains through the network, and combine these with high-resolution spin frequency measurements. First, spin-chain sensing (detailed in Section \ref{seq:SEDORception}) correlates multiple frequencies and spin-spin couplings, directly accessing the underlying network connectivity, and thus reducing ambiguity due to (potential) spectral overlap. Consider the previous example: by probing the correlation between the three frequencies $A_1, A_2$ and $A_3$ in a single measurement, we directly reveal that Spin 1 and Spin 3 are connected to the same spin at $A_2$ (Spin 2). Such a spin-chain measurement yields a correlated list of 5 frequencies: $\{ A_1, C_{12}, A_2, C_{23}, A_3\}$, characterising the 3-spin chain. Applying the same method but now with spin 4 ($A_3 \leftarrow A_4$) reveals that it is not connected to Spin 2, but couples to another spin (spin 5) that overlaps in frequency with Spin 2.  

Second, spin-chain sensing enables measuring couplings that are otherwise challenging to access, enabling exploration further into the network. Consider the case where starting from some spin (e.g. Spin 1 in Fig. \ref{fig:spin_networks}c) it is challenging to probe a part of the network, either because the couplings to Spin 1 are too weak to be observed or spectral crowding causes signals to overlap. The desired interactions  (e.g. those belonging to Spin 4 in Fig. \ref{fig:spin_networks}c) can be reached by constructing a spin chain, in which each link is formed by a strong and resolvable spin-spin interaction. The chain iteratively unlocks new spins that can be used as sensors of their own local spatial environment. 

Finally, we combine the spin-chain measurements with a correlated double-echo spectroscopy scheme that increases the resolution with which different $A_i$ are distinguished from $\sim 1/\tstar$ to $\sim 1/\ttwo$ (Fig. \ref{fig:spin_networks}d). This directly reduces spectral overlap of spin frequencies, further removing ambiguity. 

In principle, the entire network can be mapped by expanding and looping a single chain. In practice, measuring limited-size chains is sufficient. A N-spin chain measurement yields a correlated list of $N$ spin frequencies $A$, alongside $N-1$ coupling frequencies $C$, which quickly becomes uniquely identifiable, even when some spin frequencies in the network are degenerate. This allows for the merging of chains that share a common section to reconstruct the network (Methods).

\subsection{Experimental system}
We demonstrate these methods on a network of $50$ $^{13}$C spins surrounding a single NV center in diamond at 4 K. The NV electron spin is initialized and measured optically and is used as the sensor spin \cite{abobeih_atomicscale_2019}. We employ dynamical decoupling sequences to sense nuclear spins at selected frequency bands, using sequences with and without radio-frequency driving (DDRF) of the nuclear spins to ensure sensitivity in all directions from the NV (Methods) \cite{abobeih_atomicscale_2019,bradley_tenqubit_2019}. The nuclear spins are polarized via the electron spin, using global dynamical-nuclear-polarisation techniques (PulsePol sequence \cite{schwartz_robust_2018,randall_manybody_2021}), or by selective projective measurements or SWAP gates \cite{abobeih_atomicscale_2019,bradley_tenqubit_2019}.  

The $^{13}$C nuclear spin frequencies are given by $A_i = \omega_\mathrm{L} + m_s\deltah{i}$, with $\omega_\mathrm{L}$ the global Larmor frequency and $\deltah{i}$ a local shift due to the hyperfine interaction with the NV center (see for example Ref. \cite{taminiau_detection_2012} and Supplementary Note 1). Here, we neglected corrections due to the anisotropy of the hyperfine interaction, which are treated in Supplementary Note 4. The experiments are performed with the electronic spin in the $m_s=\pm 1$ states. Because, for the spins considered, $\deltah{i}$ is typically two to three orders of magnitude larger than the nuclear-nuclear couplings $C_{ij}$,  nuclear-spin flip-flop interactions are largely frozen, and Eq. \ref{eq:nuclear_hamiltonian} applies (Supplementary Note 1). 

In the NV-nuclear system, spectral crowding forms a natural challenge for determining the spin-network structure. The spin frequencies are broadened by the inhomogeneous linewidth $\sim 1 /{\tstar}$, which is mainly set by the coupling to all other nuclear spins. A limited number of nuclear spins close to the NV center are spectrally isolated (defined as: $|A_i - A_j| >   1/{\tstar} \,\, \forall\,\, j \,$), making them directly accessible with electron-nuclear gates \cite{abobeih_atomicscale_2019,bradley_tenqubit_2019}, and making pairwise measurements sufficient to map the interactions. However, the hyperfine interaction, and thus $\deltah{i}$, decreases with distance ($\sim r^{-3}$), resulting in an increasing spectral density for lower $\deltah{i}$ (larger distance). Interestingly, there exists a spectrally crowded region ($|A_i - A_j| < 1/{\tstar}$) for which nuclear spins still do not couple strongly to other spins in the same spectral region ($C_{ij} \lesssim 1/{\ttwo} \,\, \forall\,\, j \,$), for example when they are on opposite sides of the NV center. Contrary to previous work \cite{abobeih_atomicscale_2019}, the methods outlined in Section \ref{sec:spin_network_mapping} allow us to measure interactions between spins in the spectrally crowded region (see Supplementary Note 2), unlocking a part of the network that was previously not accessible.

\subsection{Spin-chain sensing} \label{seq:SEDORception}

\begin{figure*}
    \centering
    \includegraphics[width =1\textwidth]{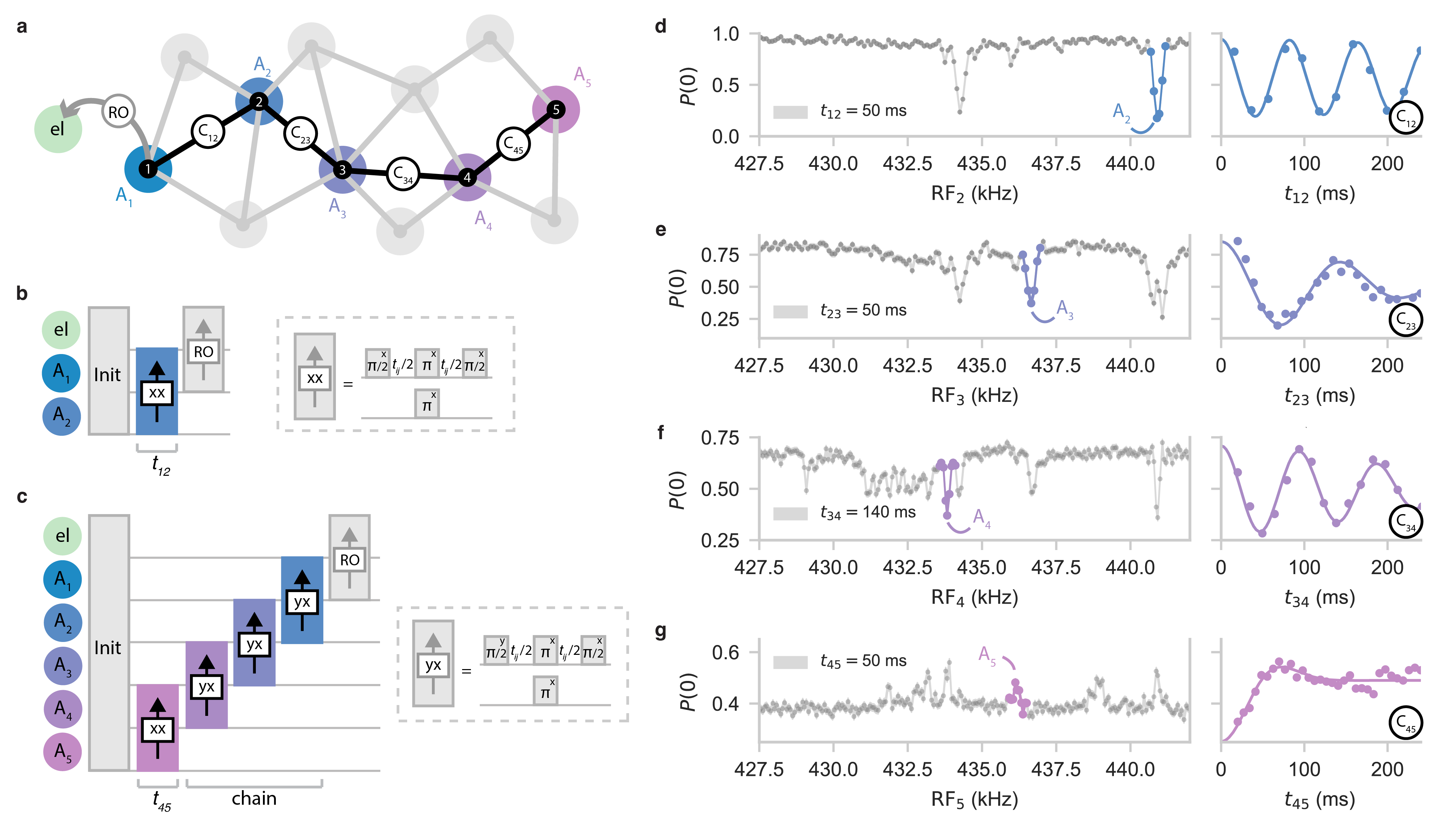}
    \caption{Sensing spin chains. a) Schematic of a $N=5$ nuclear-spin chain through different spectral regions $\{A_1,A_2,A_3,A_4,A_5\}$ (coloured disks), starting from the NV electron spin (`el'). Even though there might be multiple spins at each of the nuclear frequencies, only a single one is connected to this chain. b) Pulse sequence (Methods) for the prototypical $N=2$ sequence (SEDOR) \cite{abobeih_atomicscale_2019}. c) Pulse sequence for sensing a chain of $N=5$ nuclear spins, correlating 5 spin frequencies and 4 spin-spin interactions. In this case, the RF frequency ($\mathrm{RF}_5$) and free-evolution time ($t_{45}$) are varied to probe the connections of the spin at $A_4$ to other spins. The resulting signal is mapped back via concatenated SEDOR sequences and finally read out (`RO') through the electron spin (Methods). d-g) Experimental data, sweeping the frequency $\mathrm{RF}_N$ of the recoupling pulse (left) to detect the frequencies of spins coupled to Spin 1, and varying the free evolution time $t_{N-1,N}$ (right) to extract their coupling strengths (for $N = \{2,3,4,5\}) $. For the frequency sweeps, evolution times $t_{ij}$ are selected a-priori (annotated). Colored highlights denote the signals due to the spins in the chain and solid lines are fits to the data (see Supplementary Note 3). The signal in the bottom panel is inverted, due to the coupling $C_{34}$ being negative. }
    \label{fig:sedorception}
\end{figure*}

We experimentally demonstrate the correlated sensing of spin chains up to five nuclear spins (Fig. \ref{fig:sedorception}), by sweeping a multi-dimensional parameter space (set by 5 spin frequencies and 4 spin-spin couplings). We start by polarizing the spin network \cite{randall_manybody_2021, schwartz_robust_2018} and use the electron spin to sense a nuclear spin (Spin 1) at frequency $A_1$, which marks the start of the chain. 

First, we perform a double-resonance sensing sequence (Fig. \ref{fig:sedorception}b) consisting of a spin-echo sequence at frequency $A_1$ and an additional $\pi$-pulse at frequency $\mathrm{RF}_2$. The free evolution time $t_{12}$ is set to $50\si{\milli \second}$, to optimise sensitivity to nuclear-nuclear couplings (typically $\sim 10 \si{\hertz}$). By sweeping $\mathrm{RF}_{2}$, strong connections ($C_{1j} \gg 1/{ \ttwo}$) are revealed through dips in the coherence signal of Spin 1 (Fig. \ref{fig:sedorception}d, left). We select a connection to a spin at $\mathrm{RF}_{2} = A_2$ (Spin 2) and determine ($C_{12}$) by sweeping $t_{12}$ (Fig. \ref{fig:sedorception}d, right). 

Next, we extend the chain. To map the state of Spin 2 back to the electron sensor through Spin 1, we change the phase of the first $\tfrac{\pi}{2}$-pulse (labelled `yx') and set $t_{12} = 1/(2C_{12})$ to maximise signal transfer (see Supplementary Note 3). We then insert a double-resonance block for frequencies $\mathrm{RF}_{2} = A_2$ and $\mathrm{RF}_{3}$ in front of the sequence (Fig. \ref{fig:sedorception}c and e,left) to explore the couplings of Spin 2 to the network. This concatenating procedure can be continued to extend the chain, with up to 5 nuclear spins shown in Fig. \ref{fig:sedorception}. In general, the signal strength decreases with increasing chain length, as it is set by a combination of the degree of polarisation and decoherence ($\ttwo$  relative to $C_{ij}$) of all spins in the chain (See Supplementary Note 3). This limits the chain lengths that can be effectively used.

By mapping back the signal through the spin chain, the five spin frequencies and the 4 coupling frequencies are directly correlated: they are found to originate from the same branch of the network. As spins are now characterized by their connection to the chains, rather than by their individual, generally degenerate, frequencies (Fig. \ref{fig:spin_networks}b), they can be uniquely identified. Additionally, the chains enable measuring individual spin-spin couplings in spectrally crowded regions (Fig. \ref{fig:spin_networks}c). As an example, the expected density of spins at frequency $A_4$ is around 30 spins per kHz (Supplementary Fig. 6), making Spin 4 challenging to access directly from the electron spin. However, because Spin 3 probes only a small part of space, Spin 4 can be accessed through the chain, as demonstrated by the single-frequency oscillation in Fig. \ref{fig:sedorception}f. Another advantage over previous methods \cite{abobeih_atomicscale_2019} is that our sequences are sensitive to both the magnitude and the sign of the couplings, at the cost of requiring observable polarisation of the spins in the chain. The sign of the couplings provides additional information for reconstructing the network (Fig. \ref{fig:sedorception}g).

\subsection{High-resolution measurement of spin frequencies}\label{seq:ENES}

\begin{figure}
	\begin{center}
		\includegraphics[width=1\columnwidth]{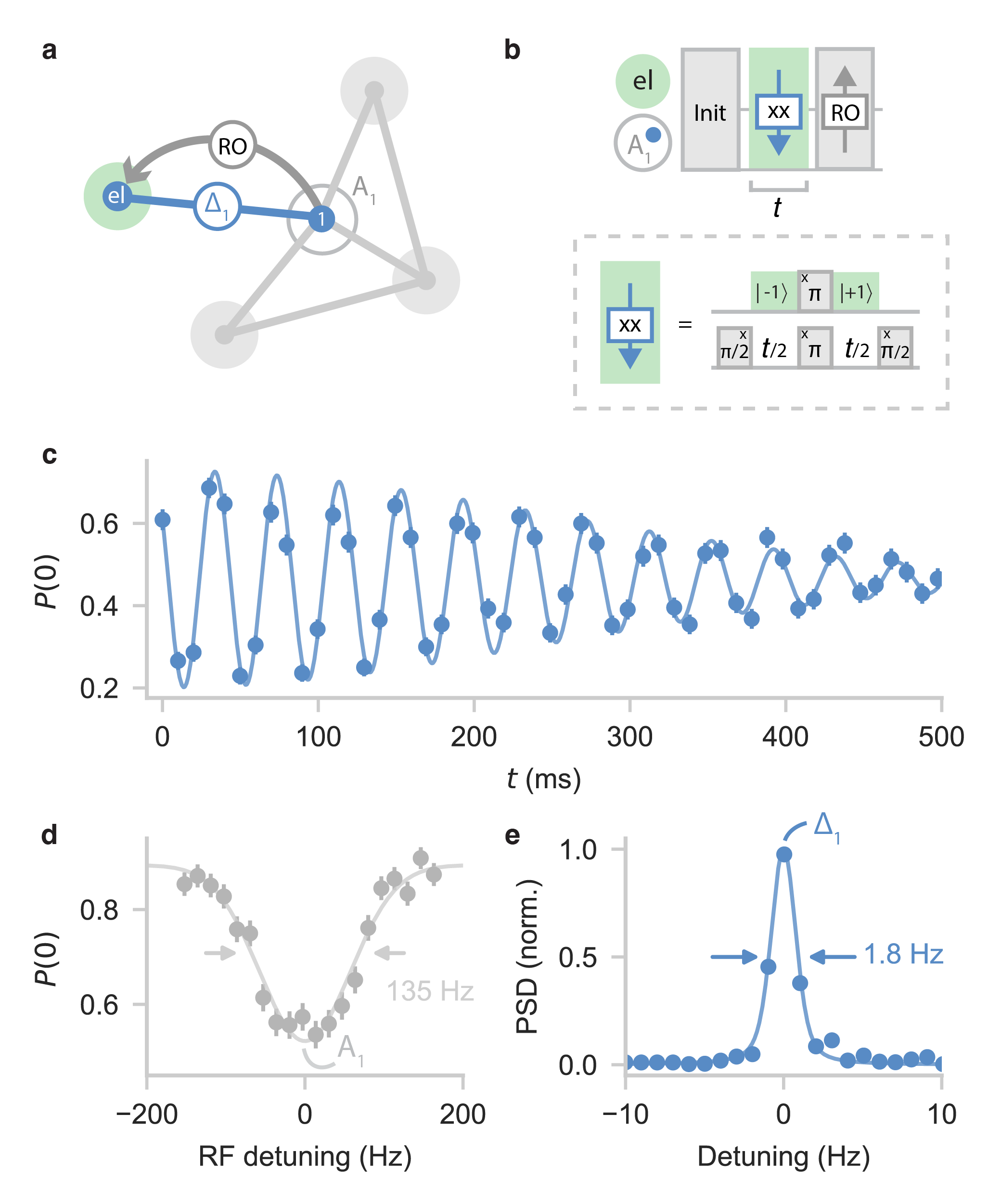}
	\end{center}
\caption{Electron-nuclear double resonance. a) The nuclear-spin frequencies $A_i$ are shifted by the hyperfine interaction with the electron spin $\deltah{i}$ (blue dotted line). Performing double resonance between the nuclear spin and the electron spin (mint green) retains this interaction while decoupling from quasi-static noise. b) Pulse sequence for measuring $\deltah{1}$ for a spin at $\approx A_1$. The nuclear spin undergoes a double resonance sequence, picking up a phase (downward arrow) from the interaction with the electron spin, whose population is synchronously transferred from the $\ket{-1}$ to the $\ket{+1}$ state. Finally, the signal is read out (denoted 'RO') via the electron spin (Methods) c) Time domain signal of $\deltah{1} = 14549.91(5)$ Hz (undersampled), with a coherence time of $T_2 = 0.36(2)$ s, fitted by a sinusoid with Gaussian decay. d) Zoom-in of spectroscopy data as in Fig. \ref{fig:sedorception}d, showing a broad resonance (135 Hz FWHM), limited by the nuclear $\tstar$-time. e) Power spectral density (PSD) of (d), showing a line width that is $\sim 75$ times improved compared to (c).}
	\label{fig:ENES}
\end{figure}

While the sensing of spin chains unlocks new parts of the network and reduces ambiguity by directly mapping the network connections, the spectral resolution for the spin frequencies ($A_i$) remains limited by the nuclear inhomogeneous dephasing time $\tstar \sim \SI{5}{\milli \second}$ \cite{bradley_tenqubit_2019}. Next, we demonstrate high-resolution ($\ttwo$-limited) measurements of the characteristic spin frequency shifts $\deltah{i}$. These frequencies provide a way to label spins, and thus further reduce ambiguity regarding which spins participate in the measured chains, particularly when a spectral region in the chain contains multiple spins (see Fig. \ref{fig:spin_networks}d). 

We isolate the interaction of nuclear spins with the electron spin through an electron-nuclear double-resonance block acting at a selected nuclear-spin frequency region. The key idea is that the frequency shift imprinted by the electron-spin sensor can be recoupled by controlling the electron spin state. We use microwave pulses that transfer the electron population from the $\ket{-1}$ to the $\ket{+1}$ state (Fig. \ref{fig:ENES}b, Methods). The nuclear spin is decoupled from quasi-static noise and the rest of the spins, extending its coherence time, while the interaction of interest ($\deltah{i}$) is retained.

Figure \ref{fig:ENES} shows an example for a nuclear spin at $A_1$, for which we measure a hyperfine shift $\deltah{1} = 14549.91(5)$ Hz and a spectral linewidth of 1.8 Hz (Fig. \ref{fig:ENES}d and e). Besides a tool to distinguish individual spins in the network with high spectral resolution, this method has the potential for improved characterisation of the hyperfine interaction in electron-nuclear spin systems. 

The observed coherence time $\ttwo =0.36(2)\si{\second}$ is slightly shorter than the bare nuclear spin-echo time $T_{2,\mathrm{SE}} = 0.62(5)$ s. This reduction is caused by a perturbitive component of the hyperfine tensor in combination with the finite magnetic field strength (see Supplementary Note 4). Flipping the electron spin between $m_s=\pm1$ changes the quantization axes of the nuclear spins, which causes a change of the nuclear-nuclear interactions \cite{abobeih_atomicscale_2019}, which is not decoupled by the spin-echo sequence (see Supplementary Fig. 1). The effect is strongest for spins near the NV center. For larger fields or for spins with weak hyperfine couplings, we expect that further resolution enhancement is possible by applying multiple refocusing pulses (see Supplementary Note 4). 

\begin{figure*}
	\begin{center}
		\includegraphics[width=1\textwidth]{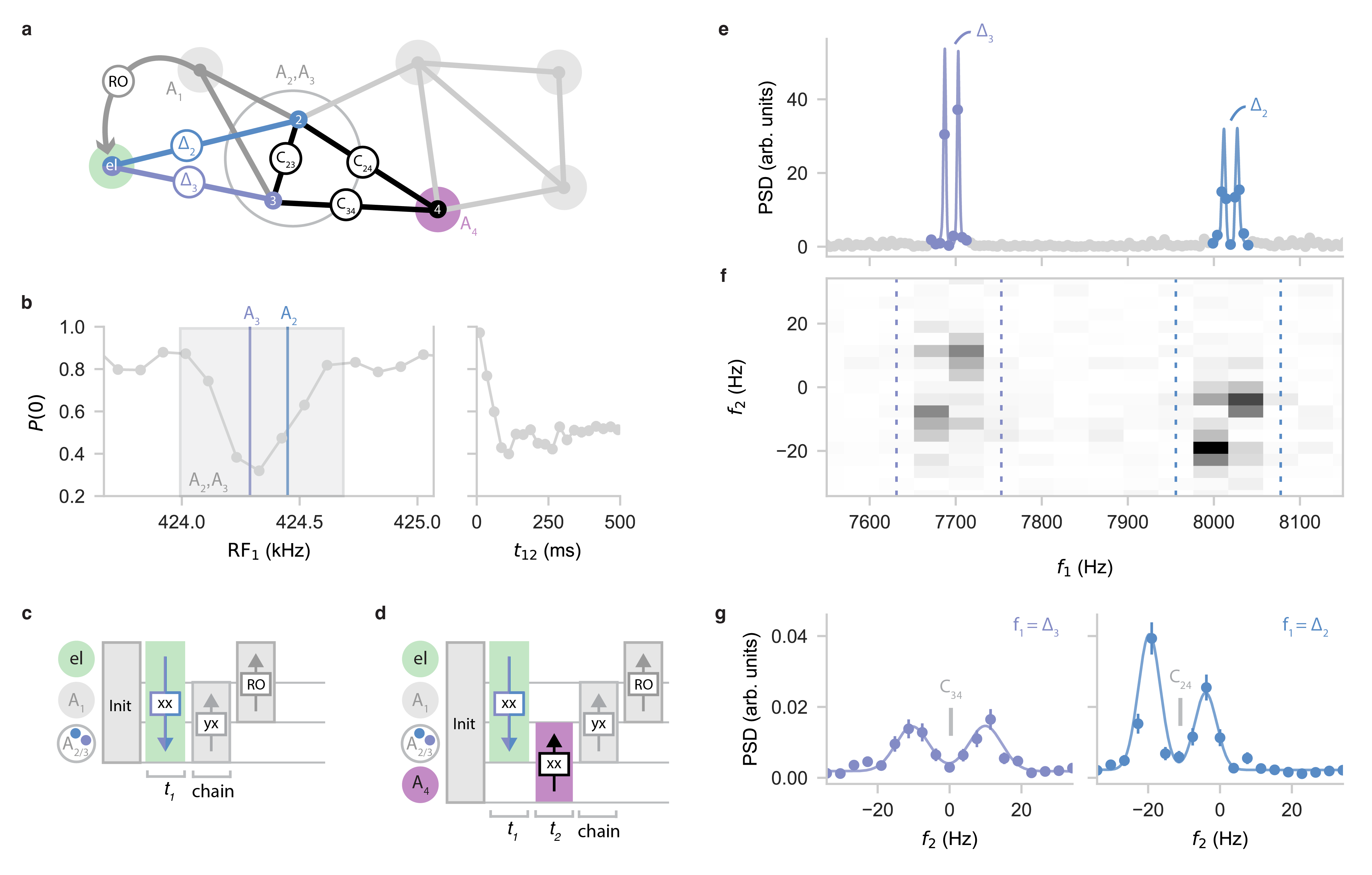}
	\end{center}
\caption{Two-dimensional spectroscopy of spectrally crowded spins. a) Schematic of the studied system, which contains two spins with overlapping frequencies  $A_2 \approx A_3$ (grey circle), with slightly different hyperfine shifts ($\deltah{2}, \deltah{3}$). Both are coupled to the spin at $A_1$, which is used for to transfer the signal to the electron spin for readout ('RO'). b) SEDOR spectroscopy (as in Fig. \ref{fig:sedorception}d) of the frequency region $A_2, A_3$, with the estimated spin frequencies indicated. Sweeping the $t_{12}$ evolution time results in a quick decay. c) Pulse sequence for the electron-nuclear double resonance sequence used in (e), where the $\deltah{i}$ are extracted by sweeping $t_1$. d) Pulse sequence combining electron-nuclear and nuclear-nuclear double resonance, used in (f). Adding a nuclear-nuclear block (pink) and sweeping both $t_1$ and $t_2$ reveals the correlation between $\deltah{i}$ and spin-spin couplings. e) Sweeping $t_1$ yields a high-resolution PSD of the $A_2, A_3$ frequency region, showing two (split) frequencies $\deltah{2}$ and $\deltah{3}$. The solid curve is a four-frequency fit to the data. f) Signal (PSD) for the two-dimensional sequence, revealing two distinct regions along the $f_1$-axis at $\deltah{2}$ and $\deltah{3}$. g) Binned line cut of (f) along the $f_2$-axis at frequencies $\deltah{2}, \deltah{3}$ (region indicated by dotted lines). The positions of the (split) peaks indicate the coupling to the spin at $A_4$ ($C_{24} = -11.8(2)$ Hz, $C_{34} = -0.2(5)$ Hz). The solid line is a fit of two Gaussians to extract the couplings.}
	\label{fig:ENES_2D}
\end{figure*}

Finally, we combine spin-chain sensing and electron-nuclear double resonance to correlate high-resolution spin frequencies ($\deltah{i}$) with specific spin-spin couplings ($C_{ij}$), even when a chain contains multiple spins with overlapping frequencies. We illustrate this scheme on a chain of spins, where two spins (2 and 3) have a similar frequency ($A_2 \approx A_3$) and both couple to $A_1$ and $A_4$ (Fig. \ref{fig:ENES_2D}a). The goal is to extract $\deltah{2}, \deltah{3}$ and the couplings to Spin 4 ($C_{24}, C_{34}$). As a reference, standard double-resonance shows a quickly decaying time-domain signal, indicating couplings to multiple spins that are spectrally unresolved (Fig. \ref{fig:ENES_2D}b).  

Figure \ref{fig:ENES_2D}c shows how the electron-nuclear double resonance sequence (mint green) is inserted in the spin-chain sequence to perform high-resolution spectroscopy of the $A_2, A_3$ frequency region. Sweeping the interaction time $t_1$ shows multiple frequencies (Fig. \ref{fig:ENES_2D}e), hinting at the existence of multiple spins with approximate frequency $A_2$. The result is consistent with two spins at frequencies $\deltah{2} = 8019.5(2)$ Hz and $\deltah{3} = 7695.2(1)$ Hz, split by an internal coupling of $C_{23} =7.6(1)$ Hz (Fig. \ref{fig:ENES_2D}a and Supplementary Fig. 2e,f). 

Next, we add a nuclear-nuclear block (pink block in Fig. \ref{fig:ENES_2D}d) and sweep both electron-nuclear ($t_1$) and nuclear-nuclear ($t_2$) double-resonance times to correlate $\deltah{2}$ and $\deltah{3}$ with nuclear-nuclear couplings $C_{24}$ and $ C_{34}$. After the $t_1$ evolution, the hyperfine shifts $\deltah{i}$ are imprinted in the z-expectation value of each spin, effectively modulating the nuclear-nuclear couplings observed in $t_2$. The 2D power spectral density (PSD) shows signals in two distinct frequency regions along the $f_1$-axis, corresponding to $\deltah{2}$ and $\deltah{3}$ (Fig. \ref{fig:ENES_2D}f). Analysing the nuclear-nuclear ($f_2$) signal at these frequencies (Fig. \ref{fig:ENES_2D}g), we find $C_{24} =-11.8(2)$ Hz and $C_{34} = -0.2(5)$ Hz. We attribute the splitting to the coupling $C_{23}$ between Spins 2 and 3 (Methods, Supplementary Fig. 2g,h). Varying $\mathrm{RF}_4$ enables the measurement of the interactions of spins 2 and 3 to other parts of the network (for example to determine $C_{12}, C_{13}$). Beyond the examples shown here, the electron-nuclear block can be inserted at specific positions in the spin-chain sequence (Fig. \ref{fig:sedorception}c) to extract $\deltah{i}$ of all spins in the chain (Supplementary Fig. 9). 

\subsection{Reconstruction of a 50-spin network} \label{seq:positioning}
\begin{figure}
	\begin{center}
		\includegraphics[width=1\columnwidth]{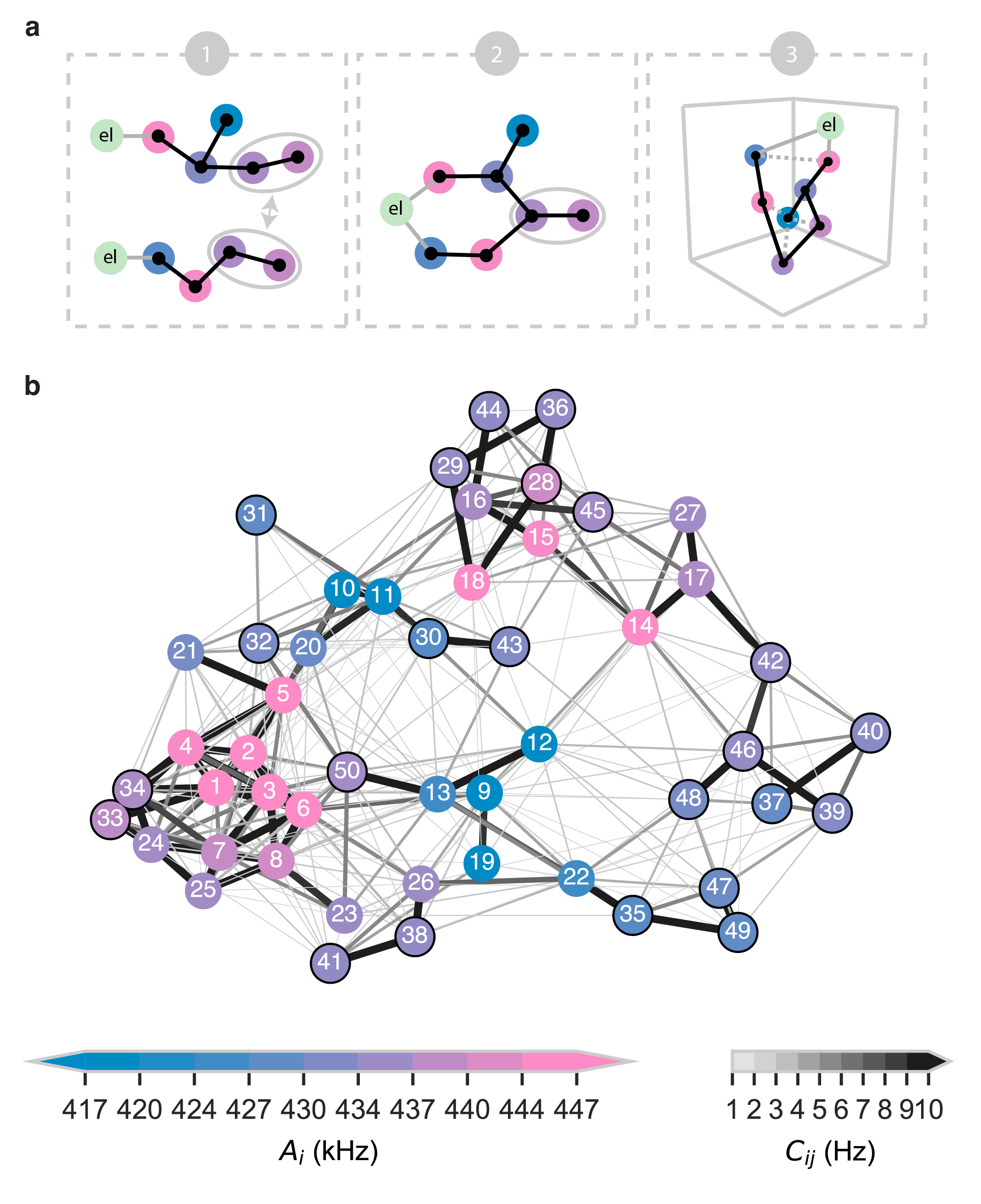}
	\end{center}
\caption{Mapping a 50-spin network. a) Schematic illustrating the procedure for mapping large networks. 1: Separate high-resolution chains through the network are measured (two example chains shown here). 2: We merge chains that share a common section of the network. 3: Optionally, an algorithm adapted from \cite{abobeih_atomicscale_2019} estimates the most likely spin positions (Methods), which predicts all unmeasured nuclear-nuclear couplings (dotted lines) and provides a validation for the assignment and merging of step 2. b) Graph of the 50-spin network mapped in this work, with edges indicating spin-spin interactions above 2 Hz and vertex colors denoting spin frequencies $A_i$. The spins are labelled according to Supplementary Table 1. Black circles indicate the 23 newly mapped spins compared to previous work \cite{abobeih_atomicscale_2019}. A 3D spatial image of the network is presented in Supplementary Fig. 3.}
	\label{fig:full_cluster}
\end{figure}

Finally, we apply these methods to map a 50-spin network. The problem resembles a graph search (Methods) \cite{gross_handbook_2004}. By identifying a number of spin chains in the system, and fusing them together based on overlapping sections, we reconstruct the connectivity (Fig. \ref{fig:full_cluster}). Limited sized chains are sufficient because the couplings are highly non-uniform, so that a few overlapping vertices and edges enable fusing chains with high confidence. We use a total of 249 measured interactions through pairwise and chained measurements. Fusing these together provides a hypothesis for the network connectivity (Fig. \ref{fig:full_cluster}b).  

To validate our solution for the network we use the additional information that the nuclear-nuclear couplings can be modeled as dipolar and attempt to reconstruct the spatial distribution of the spins. Compared to work based on pairwise measurements \cite{abobeih_atomicscale_2019}, our spin-chain measurements provide additional information on the connectivity and coupling signs, reducing the complexity of the numerical reconstruction. Additionally, we constrain the position using the measured hyperfine shift $\deltah{i}$ (Methods). Because the problem is highly overdetermined \cite{abobeih_atomicscale_2019}, the fact that a spatial solution is found that closely matches the measured frequencies and assignments validates the obtained network connectivity. Additionally, the reconstruction yields a spatial image of the spin network and predicts the remaining unmeasured 976 spin-spin interactions, most of which are weak ($< 1$Hz). An overview of the complete $50$ spin cluster, characterized by $50$ spin frequencies and 1225 spin-spin couplings can be found in Supplementary Table 1 and in Fig. \ref{fig:full_cluster}b.   

\section{Discussion}
In conclusion, we developed correlated double-resonance sensing that can map the structure of large networks of coupled spins, with high spectral resolution. We applied these methods to reconstruct a 50-spin network in the vicinity of an NV center in diamond. The methods can be applied to a variety of systems in different platforms, including electron-electron spin networks \cite{awschalom_quantum_2018,debroux_quantum_2021,sipahigil_integrated_2016,bourassa_entanglement_2020,lukin_integrated_2020,durand_broad_2021,higginbottom_optical_2022,gao_nuclear_2022,ruskuc_nuclear_2022,ungar_identification_2023}. Mapping larger spin systems might be in reach using machine-learning-enhanced protocols and sparse or adaptive sampling techniques, which can further reduce acquisition times \cite{martina_deep_2023, jung_deep_2021}. Combined with control fields \cite{randall_manybody_2021, bradley_tenqubit_2019, abobeih_faulttolerant_2022}, the methods developed here provide a basis for universal quantum control and readout of the network, which has applications in quantum simulations of many-body physics \cite{randall_manybody_2021}. Furthermore, the precise characterisation of a 50-spin network provides new opportunities for optimizing quantum control gates in spin qubit registers \cite{bradley_tenqubit_2019, abobeih_faulttolerant_2022, xie_99_2023}, for testing theoretical predictions for defect spin systems \cite{nizovtsev_nonflipping_2018}, and for studying coherence of solid-state spins on the microscopic level, including quantitative tests of open quantum systems and approximation of the central spin model \cite{yang_quantum_2016}. Finally, these results might inspire high-resolution nano-MRI of quantum materials and biologically relevant samples outside the host crystal. \\

% \clearpage
% \newpage
% \onecolumngrid
\section{Methods}
\subsection{Sample and setup}
All experiments are performed on a naturally occurring NV center at a temperature of 3.7K (Montana S50 Cryostation), using a home-built confocal microscopy setup. The diamond sample was homoepitaxially grown using chemical vapor deposition and cleaved along the $\langle 111 \rangle$ crystal direction (Element Six). The sample has a natural abundance of $^{13}$C (1.1 \%). The NV center has been selected on the absence of couplings to $^{13}$C stronger than $\approx 500$ kHz. No selection was made on other properties of the $^{13}$C nuclei distribution. A solid immersion lens (SIL) that enhances photon collection efficiency is fabricated around the NV center. A gold stripline is deposited close to the edge of the SIL for applying microwave (MW) and radio-frequency (RF) pulses. An external magnetic field of $B_z=403.553$ G is applied along the symmetry axis of the NV center, using a (temperature-stabilized) permanent neodymium magnet mounted on a piezo stage outside the cryostat \cite{bradley_tenqubit_2019}. The field is aligned to within 0.1 degrees using a thermal echo sequence \cite{abobeih_atomicscale_2019}. 

\subsection{Electron and nuclear spins}
The sample was previously characterized in Abobeih et al. \cite{abobeih_atomicscale_2019} and the 27 nuclear spins imaged in that work are a subset of the 50 nuclear-spin network presented here. The NV electron spin has a dephasing time of $\tstar = \SI{4.9(2)}{\micro \second}$, a Hahn spin echo time of $\ttwo = 1.182(5)$ ms, and a relaxation time of $T_1>1$ hr \cite{abobeih_atomicscale_2019}. The spin state is initialized via spin-pumping and read out in a single shot through spin-selective resonant excitation, with fidelities $F_0 = 89.3(2)$ ($F_1 = 98.2(1)$) for the the $m_s = 0$ ($m_s = -1$) state, resulting in an average fidelity of $F_{\mathrm{avg}}=0.938(2)$. The readout is corrected for these numbers to obtain a best estimate of the electronic spin state. The nuclear spins have typical dephasing times of $\ttwo = 5 - 10$ ms and Hahn echo $T_{2,n}$ up to $0.77(4)$ s \cite{bradley_tenqubit_2019}. $\ttwo$-times for spins with frequencies closer to the nuclear Larmor frequency ($\Delta_i \lesssim 5$ kHz) typically decrease to below $100$ ms (see e.g. Fig. \ref{fig:sedorception}g, right panel), as the spin echo simultaneously drives other nuclear spins at these frequencies which are re-coupled to the target (instantaneous diffusion).

\subsection{Pulse sequences}
We drive the electronic $m_s = 0 \leftrightarrow m_s = -1$ ($m_s = 0 \leftrightarrow m_s = +1$) spin transitions at 1.746666 (4.008650) GHz with Hermite-shaped pulses. For transferring the electron population from the $m_s = -1$ to the $m_s = +1$ state (Fig. \ref{fig:ENES} and \ref{fig:ENES_2D}), we apply two consecutive $\pi$-pulses at the two MW transitions, spaced by a waiting time of 3 $\mu$s. For all experiments, we apply RF pulses with an error-function envelope in the frequency range $400 - 500$ kHz. Details on the electronics to generate these pulses can be found in Ref. \cite{randall_manybody_2021}. 

For most experiments described in this work, the measurable signal is dependent on the degree of nuclear spin polarisation. We use a dynamical nuclear polarisation sequence, PulsePol,to transfer polarisation from the electron spin to the nuclear spin bath \cite{schwartz_robust_2018, randall_manybody_2021}. The number of repetitions of the sequence is dependent on the specific polarisation dynamics of the spins being used in the given experiment, but ranges from 500-10000. The PulsePol sequence is indicated by the `Init' block in the sequence schematics. All double resonance sequences follow the convention illustrated in the dotted boxes in Fig. \ref{fig:sedorception}b,c and Fig. \ref{fig:ENES}b, where the horizontal grey lines denote different RF frequencies and the top line the electronic MW frequency. The two letters in the double resonance blocks (`xx' or `yx') denote the rotation axes of the first and final $\pi / 2$-pulses. The $\pi$-pulses (along the x-axis) are applied sequentially (following Ref. \cite{abobeih_atomicscale_2019}). The lengths of all RF pulses are taken into account for calculating the total evolution time. Nuclear spins are read out via the electron by phase-sensitive (`yx') dynamical decoupling; DD or DDRF sequences \cite{bradley_tenqubit_2019}, indicated by the `RO'-marked block in the sequence schematics. Typically, the spin that is read out with the electron is reinitialised via a SWAP gate before the final SEDOR block in order to maximise its polarisation. However, all experiments presented here can be performed by using just the DNP initialisation, albeit with a slightly lower signal to noise ratio.

\subsection{2D spectroscopy experiments}
For the 2D measurement we concatenate an electron-nuclear double resonance with a nuclear-nuclear SEDOR.  For every $t_1$-point, we acquire 20 $t_2$ points, ranging from $10$ to $260$ ms. The final $\pi/2$-pulse of the electron double resonance and the first of the SEDOR are not executed, as they can be compiled away. To correct for any slow magnetic field drifts that lead to miscalibration of the two-qubit gate used for read-out, causing a small offset in the measured signal, we set our signal baseline to the mean of the final five points ($\approx 200-260$ ms), where we expect the signal to be mostly decayed. Note that these field drifts do not affect any of the double resonance blocks in which the quantities to be measured are encoded (due to the spin-echo). 

Both the 1D (Fig. \ref{fig:ENES_2D}d) and 2D (Fig. \ref{fig:ENES_2D}e) signals are undersampled to reduce the required bandwidth. To extract $\Delta A_2, \Delta A_3$, we fit a sum of cosines to the time domain signal of Fig. \ref{fig:ENES_2D}d. To extract the frequencies along the $f_2$-axis, which encode the nuclear-nuclear couplings ($C_{24},C_{34}$), we take an (extended) line-cut at $f_1 = \Delta A_2$ and $f_1 = \Delta A_3$. To increase the signal, we sum over the four bins indicated by the dotted lines. We fit two independent Gaussians to the $f_2$-data to extract $C_{24}$ and $C_{34}$. We find splittings of $7.8(2)$ Hz and $10.2(5)$ Hz, respectively, whose deviation with respect to measurements in Fig. \ref{fig:ENES_2D}d is unexplained. The skewed configuration of the two peaks (lower left, upper right) is a result of the correlation of the neighbouring spin state between the $t_1$ and $t_2$ evolution times. The different ratio of signal amplitudes belonging to Spin 2 and Spin 3, between the 1D and 2D electron-nuclear measurements are due to using different settings for the chained readout (evolution time, RF power). As we are only interested in extracting frequencies, we can tolerate such deviations.

Supplementary Figure 2 shows numerical simulations of the experiments presented in Fig. \ref{fig:ENES_2D}. These are generated by evaluating the Hamiltonian in Supplementary Eq. 3, taking into account the two spins at $A_{2},A_{3}$, the spin at $A_1$ and the electron spin.

\subsection{Network reconstruction}
Here, we outline a general procedure for mapping the network by performing specific spin-chain and high-resolution $\deltah{i}$ measurements. The mapping-problem resembles a graph search, with the NV electron spin used as root \cite{gross_handbook_2004}. We base the protocol on a breadth-first like search, which yields a spanning tree as output, completely characterising the network. The following pseudocode describes the protocol:
\newline 
\begin{algorithmic}
\State $Input:$ physical spin network, initial vertex $el$
\State $Output:$ breadth-first tree $T$ from root $el$
\State $V_0  = \{el\}$ \Comment{Make $el$ the root of $T$, $V_i$ denotes the set of vertices at distance $i$}
\State $i = 0$
\While{$V_i \neq \emptyset $} \Comment{Continue until network is exhausted}
    \For{each vertex $v \in V_i$}
        \For{each frequency $f$}
            \State C, singlecoupling = MeasureCoupling($v,f$) \Comment{Returns coupling $C$ between vertex $v$ and frequency $f$}
            \If{singlecoupling} \Comment{Checks if MeasureCoupling returned a single, resolvable coupling}
                \State create vertex $w$
                \State $A_w = f$
                \State $C_{vw} = C$ 
                \State unique, duplicate = CheckVertex($w,T$) \Comment{Checks if $w$ was already mapped in $T$}
                \If{unique} \Comment{$w$ was not yet mapped}
                    \State add $w$ to $V_{i+1}$ in $T$ \Comment{$w$ is added to $T$ as a new vertex}
                    % \State add $(v,w) = C$ to $T$
                \EndIf
                \If{$\mathbf{not}$ unique $\mathbf{and}$ duplicate == $k$} \Comment{$w$ is the same vertex as $k$ in $T$}
                    \State add $C_{vk} = C_{vw}$ in $T$ \Comment{The measured coupling is assigned to $k$}
                    \State delete $w$
                \EndIf
                \If{$\mathbf{not}$ unique $\mathbf{and}$ duplicate == \texttt{None}} \Comment{Undecided if spin was mapped}
                    \State delete $w$ \Comment{$w$ is not added to $T$}
                \EndIf
            \EndIf
        \EndFor
    \EndFor
    \State $i = i + 1$
\EndWhile
\newline
\end{algorithmic}
New vertices that are detected by chained measurements are iteratively added, once we verify that a vertex was not characterised before (i.e. has a duplicate in the spanning tree $T$). The function MeasureCoupling($v,f$) performs a spin-echo double resonance sequence between vertex $v$ and a frequency $f$, (a spin chain of length $i-1$ is used to access $v$) and checks whether a single, resolvable coupling is present (stored in the boolean variable `singlecoupling'). In the case that $v$ is the electron spin ($el$) an electron-nuclear DD(RF) sequence is performed \cite{taminiau_detection_2012,bradley_tenqubit_2019}. The function CheckVertex($w,T$) instructs the experimenter to perform a number of spin-chain and electron-nuclear double resonance measurements, comparing the vertex $w$ and its position in the network with that of the (possibly duplicate) vertex $k$ (see Supplementary Note 5). If one of these measurements is not consistent with our knowledge of $k$, we conclude $w$ is a unique vertex and add it to $T$. If all measurements coincide with our knowledge of $k$, we conclude it is the same vertex and merge $w$ and $k$. If the CheckVertex($w,T$) is inconclusive (e.g. due to limited measurement resolution), we do not add $w$ to $T$. Note that the measurement resolution, determined by the nuclear $\ttwo$-time, is expected to decrease for spins further away from the NV center (See Supplementary Note 2). This eventually limits the number of unique spins that can be identified and added to the network map.

The platform-independent procedure outlined above can be complemented by logic based on the 3D spatial structure of the system \cite{abobeih_atomicscale_2019}. For example, when the CheckVertex($w,T$) function is inconclusive, one can sometimes still conclude that $w$ must be unique (or vice versa equal to $k$), based on the restricted number of possible physical positions of these two spins in 3D space \cite{abobeih_atomicscale_2019}. In practice, we alternate the graph search procedure with calls to a positioning algorithm \cite{abobeih_atomicscale_2019}, which continuously checks whether the spanning tree $T$ is physical and aides in the identification of possible duplicates.

\subsection{3D spatial image}
For the 3D reconstruction of the network, we use the positioning algorithm developed in Ref. \cite{abobeih_atomicscale_2019}. To limit the experimental time we re-use the data of Ref. \cite{abobeih_atomicscale_2019} and add the new measurements to it in an iterative way. We set the tolerance for the difference between measured and calculated couplings to $1$ Hz. Although we only measure the new spin-spin couplings and chains when the electron is in the $m_s=-1$ state, we can assume this is within tolerance to the average value of the coupling if the perpendicular hyperfine component is small ($<10$kHz) \cite{abobeih_atomicscale_2019}. The spin positions are restricted by the diamond lattice. Spins that belong to the same chain are always added in the same iteration and up to 10000 possible configurations are kept. Chains starting from different parts of the known cluster can be positioned in a parallel fashion if they share no spins, reducing computational time. For spins that are relatively far away from the NV, we also make use of the interaction with the electron spin, approximating the hyperfine shift $\deltah{i}$ to be of dipolar form within a tolerance of 1 kHz (neglecting the Fermi contact term \cite{nizovtsev_nonflipping_2018}). For those cases, we model the electron spin as a point dipole with origin at the center of mass, as computed by density functional theory \cite{nizovtsev_nonflipping_2018}. If multiple solutions are found, we report the standard deviation of the possible solutions as a measure of the spatial uncertainty (see Supplementary Table 1).

\subsection{Error model and fitting}
Confidence intervals assume the measurement of the electron state is limited by photon shot-noise. The shot-noise-limited model is propagated in an absolute sense, meaning the uncertainty on fit parameters is not rescaled to match the sample variance of the residuals after the fit. For all quoted numbers, the number between brackets indicates one standard deviation or error indicated by the fitting procedure. We calculate the error on the PSD according to Ref. \cite{hoyng_error_1976}, assuming normally distributed errors.

% \clearpage
\section{Data availability}
All data underlying the study are available on the open 4TU data server under accession code: https://doi.org/10.4121/aba1cc84-0aea-4cdc-93ca-68b0db38bd81.v1. 

\section{Code availability}
Code used to operate the experiments is available on request.

\section{References}
% \bibliographystyle{naturemag}
% \bibliography{references}

\section{Acknowledgements} 
We thank V. V. Dobrovitski for useful discussions, and H.P. Bartling and S.J.H. Loenen for experimental assistance. This work is part of the research programme NWA-ORC (NWA.1160.18.208 and NWA.1292.19.194), (partly) financed by the Dutch Research Council (NWO). This work was supported by the Dutch National Growth Fund (NGF), as part of the Quantum Delta NL programme. This work was supported by the Netherlands Organisation for Scientific Research (NWO/OCW) through a Vidi grant. This project has received funding from the European Research Council (ERC) under the European Union’s Horizon 2020 research and innovation programme (grant agreement No. 852410). This work was supported by the Netherlands Organization for Scientific Research (NWO/OCW), as part of the Quantum Software Consortium Program under Project 024.003.037/3368. S.A.B. and L.C.B. acknowledge support from the National Science Foundation under grant ECCS-1842655, and from the Institute of International Education Graduate International Research Experiences (IIE-GIRE) Scholarship. S.A.B. acknowledges support from an IBM PhD Fellowship.

\section{Author contributions}
GLvdS, DK and THT devised the experiments. GLvdS performed the experiments and collected the data. CEB, JR, SAB, LCB, MHA and THT performed and analyzed preliminary experiments. GLvdS, DK, CEB and JR prepared the experimental apparatus. GLvdS, DK and THT analyzed the data. MM and DJT grew the diamond sample. GLvdS and THT wrote the manuscript with input from all authors. THT supervised the project.

\section{Competing Interests}
T.H. Taminiau, G.L. van de Stolpe, C.E. Bradley, J. Randall and D.P. Kwiatkowski declare competing interest in the form of a Dutch patent application. Patent applicant: Technische Universiteit Delft. Name of inventor(s): Taminiau, Tim Hugo; van de Stolpe, Guido Luuk; Bradley, Conor Eliot; Randall, Joe; Kwiatkowski, Damian Patryk. Application number: NL2035279. Status of application: filed, waiting for search report. Covered aspects: Full content of manuscript.

M.H. Abobeih, S.A. Breitweiser, L.C. Basset, M. Markham and D.J. Twitchen declare no competing interests.

\newpage
\input{supplement}

\end{document}

%% file: supplement.tex
\onecolumngrid

% \section*{Supplementary Information}
\clearpage
\widetext
\begin{center}
\textbf{\large Supplementary Information for \\``Mapping a 50-spin-qubit network through correlated sensing''}
\end{center} 

\renewcommand{\theequation}{\arabic{equation}}
\setcounter{equation}{0}

\renewcommand{\figurename}{Supplementary Fig.}
\setcounter{figure}{0}

\renewcommand{\tablename}{Supplementary Table}
\setcounter{table}{0}

\def\arraystretch{1.4}
\setlength\tabcolsep{9pt}
\begin{longtable}{c|ll|rr|rrr} 
\caption{Information on the nuclear spins mapped in this work. A dash denotes that no $\deltah{i}$ signal could be obtained due to low polarisation, coherence, or readout contrast. Value in parentheses denotes standard deviation on the last digit. Spins C1-C27 were previously characterised in Ref. \cite{abobeih_atomicscale_2019} and the used labels are consistent with their work. These data are also available online at: \textit{https://doi.org/10.4121/aba1cc84-0aea-4cdc-93ca-68b0db38bd81.v1}}\\ 
\toprule
Label & Initialisation & Readout & $A_i$ (kHz) & $\Delta_i$ (Hz) & x (\AA) & y (\AA) & z (\AA) \\
\midrule
\endfirsthead
\caption[]{Information on the nuclear spins mapped in this work. A dash denotes that no $\deltah{i}$ signal could be obtained due to low polarisation, coherence, or readout contrast. Value in parentheses denotes standard deviation on the last digit. Spins C1-C27 were previously characterised in Ref. \cite{abobeih_atomicscale_2019} and the used labels are consistent with their work. These data are also available online at: \textit{https://doi.org/10.4121/aba1cc84-0aea-4cdc-93ca-68b0db38bd81.v1}} \\
\toprule
Paper label & Initialisation & Readout & $A_i$ (kHz) & $\Delta_i$ (Hz) & x (\AA) & y (\AA) & z (\AA) \\
\midrule
\endhead
\midrule
\multicolumn{8}{r}{{Continued on next page}} \\
\midrule
\endfoot

\bottomrule
\endlastfoot

         C1 &       PulsePol & Chain-1 &   452.83(2) &     -20840.2(6) &     0.0 &     0.0 &     0.0 \\
         C2 &       PulsePol & Chain-1 &   455.37(2) &     -22939.6(1) &    2.52 &    2.91 &   -0.51 \\
         C3 &       PulsePol & Chain-1 &   463.27(2) &     -31257.8(1) &    3.78 &    0.73 &   -0.51 \\
         C4 &       PulsePol & Chain-1 &   446.23(4) &     -14056.7(2) &   -1.26 &    2.18 &     0.0 \\
         C5 &           SWAP &  Direct &  447.234(1) &       -11291(3) &     0.0 &    4.37 &   -6.18 \\
         C6 &           SWAP &  Direct &  480.625(1) &     -48488.0(8) &    5.04 &   -1.46 &   -2.06 \\
         C7 &       PulsePol & Chain-1 &  440.288(6) &      -8332.8(1) &    5.04 &   -1.46 &    5.66 \\
         C8 &       PulsePol & Chain-1 &   441.77(1) &      -9803.6(8) &    7.57 &    1.46 &     3.6 \\
         C9 &           SWAP &  Direct &  218.828(1) &     213147.2(2) &    7.57 &   -4.37 &  -10.81 \\
        C10 &           SWAP &  Direct &  414.407(1) &      17643.3(4) &     0.0 &    8.74 &  -12.36 \\
        C11 &           SWAP &  Direct &  417.523(4) &     14549.91(4) &    6.31 &    9.46 &  -12.87 \\
        C12 &           SWAP &  Direct &  413.477(1) &      20546.7(3) &   11.35 &    0.73 &  -14.42 \\
        C13 &       PulsePol & Chain-1 &  424.449(1) &       8017.1(2) &   12.61 &    2.91 &   -6.69 \\
        C14 &           SWAP &  Direct &  451.802(1) &     -19760.5(3) &    5.04 &   -2.91 &  -22.65 \\
        C15 &       PulsePol & Chain-1 &   446.01(5) &     -13958.0(3) &    1.26 &    3.64 &  -22.65 \\
        C16 &       PulsePol & Chain-1 &   436.67(5) &      -4647.8(1) &    2.52 &    8.74 &  -23.17 \\
        C17 &       PulsePol & Chain-1 &   437.61(1) &      -5682.1(1) &    6.31 &   -2.18 &  -29.34 \\
        C18 &           SWAP &  Direct &   469.02(1) &     -36184.3(2) &     0.0 &   -1.46 &  -19.05 \\
        C19 &           SWAP &  Direct &  408.317(1) &     24219.15(8) &    3.78 &   -9.46 &   -8.75 \\
        C20 &       PulsePol & Chain-1 &  429.403(4) &       2692.5(5) &    3.78 &   10.92 &   -4.63 \\
        C21 &       PulsePol & Chain-1 &  430.937(3) &       1214.8(4) &   -5.04 &    5.82 &   -4.12 \\
        C22 &       PulsePol & Chain-1 &  424.289(3) &      7696.07(9) &   16.39 &   -3.64 &   -8.24 \\
        C23 &       PulsePol & Chain-1 &  435.143(7) &      -3195.6(1) &   13.88 &    0.73 &    5.66 \\
        C24 &       PulsePol & Chain-2 &  436.183(3) &               - &    1.26 &   -0.73 &    9.78 \\
        C25 &       PulsePol & Chain-2 &  435.829(5) &               - &    7.57 &    1.46 &    9.78 \\
        C26 &       PulsePol & Chain-1 &  435.547(2) &               - &   12.61 &   -5.82 &   -0.51 \\
        C27 &       PulsePol & Chain-1 &   435.99(3) &      -3935.9(2) &    1.26 &   -3.64 &   -31.4 \\
        C28 &       PulsePol & Chain-1 &    440.9(1) &     -8915.47(3) &   -1.26 &    2.18 &  -24.71 \\
        C29 &       PulsePol & Chain-1 &    434.3(1) &      -2185.7(1) &   -6.31 &    0.72 &  -19.05 \\
        C30 &       PulsePol & Chain-1 &    427.1(1) &      4871.11(4) &   12.62 &   10.19 &  -14.93 \\
        C31 &       PulsePol & Chain-1 &    428.3(1) &               - &   11(4) &   15(4) &  -11(4) \\
        C32 &       PulsePol & Chain-1 &    431.6(1) &               - &    6(3) &   12(2) &   -3(9) \\
        C33 &       PulsePol & Chain-1 &    439.0(1) &               - &   -2.52 &   -1.46 &    4.12 \\
        C34 &       PulsePol & Chain-1 &    437.3(1) &               - &   -2.52 &    -0.0 &    6.18 \\
        C35 &       PulsePol & Chain-1 &    427.4(1) &      4591.33(4) &   20(3) &   -1(7) &   -8(1) \\
        C36 &       PulsePol & Chain-1 &    434.4(1) &      -2214.2(8) &   -8(7) &   -0(7) &  -23(9) \\
        C37 &       PulsePol & Chain-1 &    429.1(1) &       2899.5(2) &   13(2) &  -12(1) &  -18(1) \\
        C38 &       PulsePol & Chain-1 &    434.0(1) &               - &   12.61 &   -8.73 &     0.0 \\
        C39 &       PulsePol & Chain-2 &    432.5(1) &         -450(5) &   16(5) &   -8(5) &  -24(4) \\
        C40 &       PulsePol & Chain-2 &    433.3(1) &        -1173(5) &   11(3) &  -14(3) &  -26(4) \\
        C41 &       PulsePol & Chain-2 &    434.1(1) &      -2189.3(6) &    8(4) &  -12(4) &    2(6) \\
        C42 &       PulsePol & Chain-2 &    434.8(1) &               - &   10.09 &   -7.28 &  -28.83 \\
        C43 &       PulsePol & Chain-2 &    432.2(1) &       -270.1(6) &   11(3) &    7(3) &  -19(3) \\
        C44 &       PulsePol & Chain-2 &    433.9(1) &      -1882.8(6) &    -0.0 &   11.65 &  -26.77 \\
        C45 &       PulsePol & Chain-3 &    436.2(1) &        -4174(1) &     6.3 &    6.56 &  -29.35 \\
        C46 &       PulsePol & Chain-3 &    434.8(1) &               - &   14(4) &   -5(4) &  -24(5) \\
        C47 &       PulsePol & Chain-2 &    429.4(1) &       2587.8(3) &   21(4) &   -7(7) &  -14(7) \\
        C48 &       PulsePol & Chain-2 &    431.0(1) &               - &   16(4) &   -4(4) &  -21(9) \\
        C49 &       PulsePol & Chain-2 &    428.3(1) &       3744.4(2) &   23(4) &   -4(8) &  -12(3) \\
        C50 &       PulsePol & Chain-2 &    436.2(1) &      -4227.0(4) &   11(7) &    5(7) &   -2(9) 
\label{tab:spin_overview}
\end{longtable}
\bigskip

% \clearpage
\begin{table}[ht]
    \def\arraystretch{1.4}
    \setlength\tabcolsep{9pt}
    \centering
    \caption{Spin labels of spins featuring in the experiments in the main text.}
    \begin{tabular}{l|c|c}
    \toprule
    Figure  & Spin number &  Spin label \\
    \midrule
    Fig. 2                      & Spin 1 & C18 \\
                                & Spin 2 & C28 \\
                                & Spin 3 & C16 \\
                                & Spin 4 & C44 \\
                                & Spin 5 & C45 \\
    Fig. 3                      & Spin 1 & C11 \\
    Fig. 4                      & Spin 1 & C12 \\                                        & Spin 2 & C13 \\
                                & Spin 3 & C22 \\   
    \end{tabular}

    \label{tab:my_label}
\end{table}

\clearpage

\begin{figure}[h]
	\begin{center}
		\includegraphics[width=0.88\columnwidth]{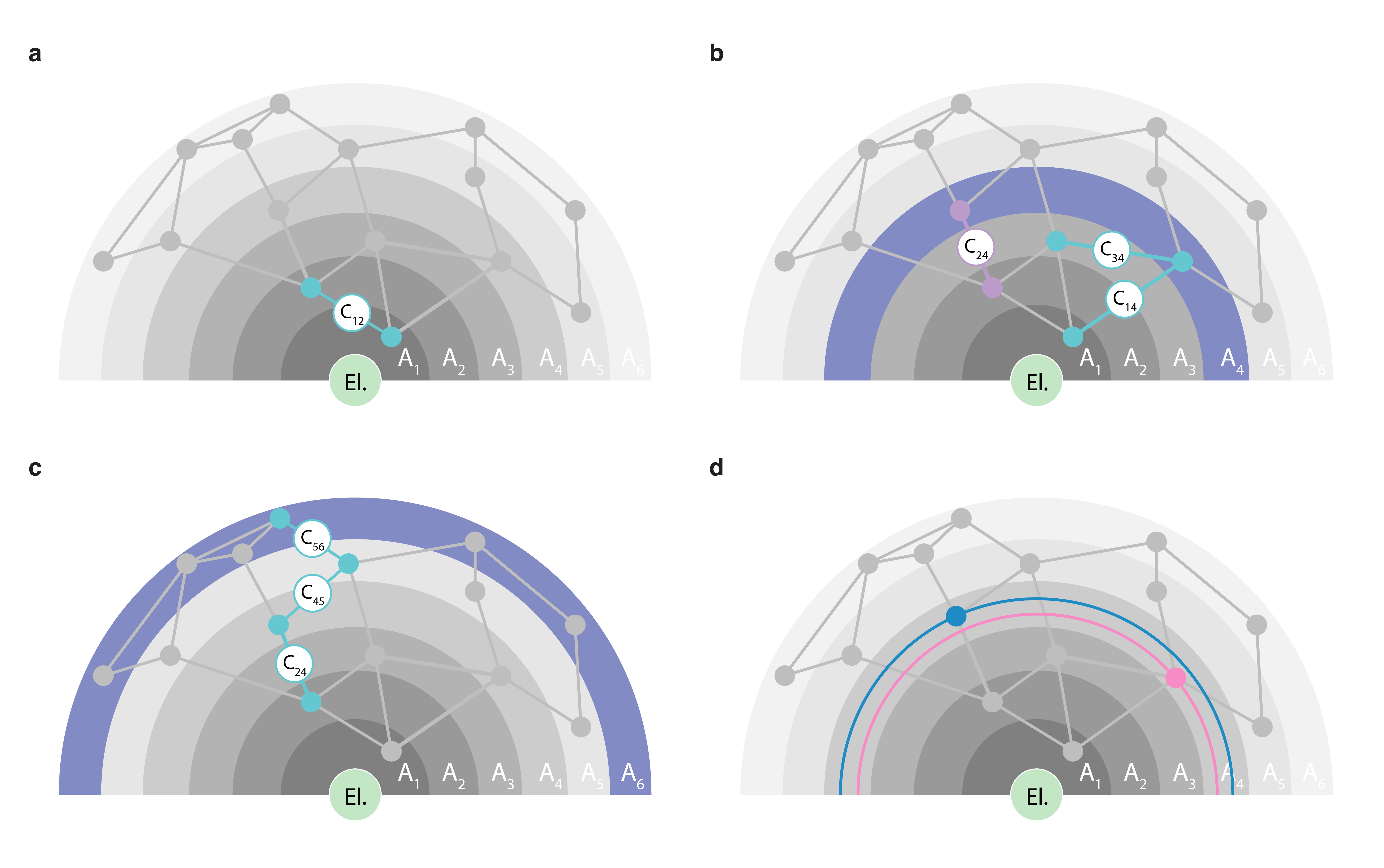}
	\end{center}
\caption{\emph{Mapping complex spin networks.} a) Spatial representation of Fig. 1, specific to the electron-nuclear system. Nuclear spins (dots) are connected by lines denoting observable couplings ($C_{ij} > 1/{\ttwo}$). The bands indicate spin frequencies $A_i$, shifted by the hyperfine interaction $\deltah{i}$ with the electron spin (mint green). This interaction diminishes with distance from the NV center, leading to spectral crowding (multiple spins per band). For simplicity we do not visualize the angular dependence of the hyperfine interaction. Frequencies $A_1$, $A_2$ and $A_3$ contain only a single spin, allowing for a direct readout with the electron spin. Additionally, the measured couplings between these frequencies (e.g. $C_{12}$) can be assigned unambiguously from pairwise measurements (see Fig. 1a). b) When multiple spins occupy the same frequency band (e.g. at $A_4$), spin-chain measurements resolve ambiguity by retrieving the connectivity of the network (see also Fig. 1b). c) Spins in spectrally crowded areas (e.g. at $A_{6}$, see also Supplementary Fig. 1), can still be accessed via a spin-chain, starting from a directly accessible spin (e.g. at $A_2$)  (see Fig. 1c). d) High-resolution measurements of $\deltah{i}$ (indicated by narrow pink and blue bands) allow for directly distinguishing multiple spins in a single frequency band (e.g. at $A_4$)  (see Fig. 1d). Note that we use different frequency labels compared to Fig. 1.}
    \label{fig_ext:spectral_regions}
\end{figure}

\begin{figure}
	\begin{center}
		\includegraphics[width=1\columnwidth]{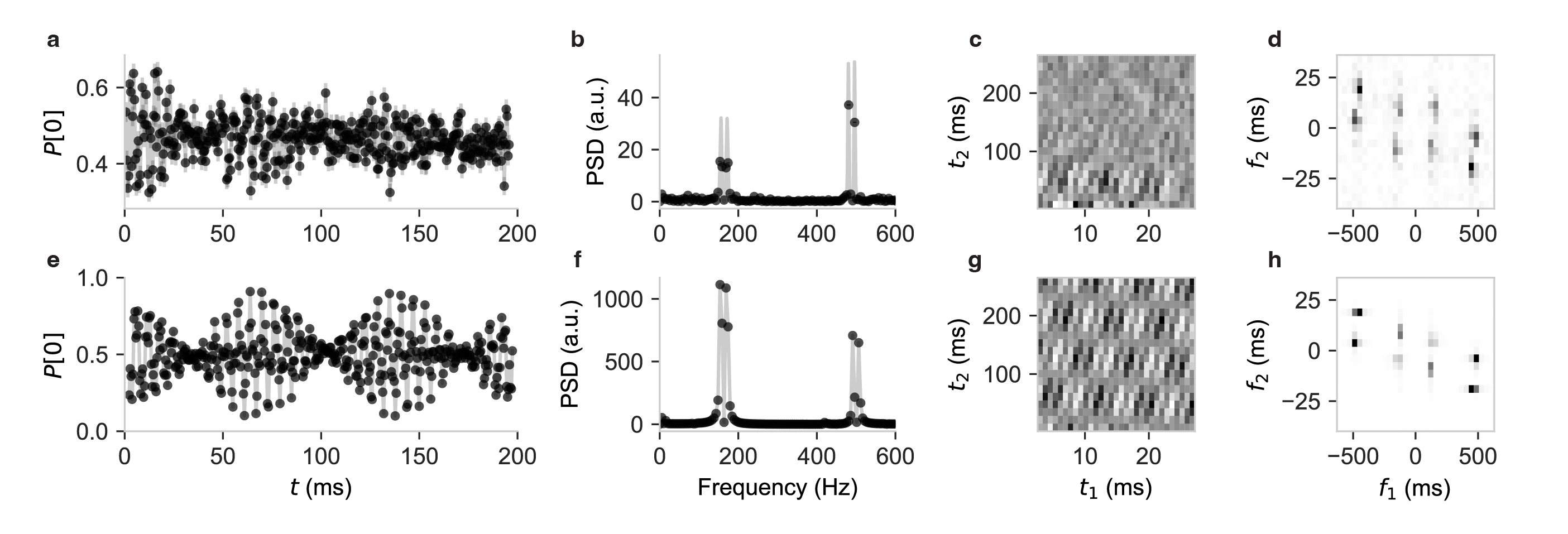}
	\end{center}
\caption{\emph{Comparison to numerical simulations}. a) Time domain data of the experiment in Fig. 4d and corresponding PSD (b). c) Time domain data of the experiment in Fig. 4e and corresponding 2D PSD (d), showing both the positive and negative $f_1$-axis. e-h) Numerical simulations of the experimental data in (a-d). Spin parameters are based on the spin positions. The simulations show good agreement with experiment (up to nuclear decoherence effects), reconfirming the characterisation of the system and the interpretation of the spectroscopy data.}
    \label{fig_ext:numerical_simulations}
\end{figure}

\begin{figure}
	\begin{center}
		\includegraphics[width=0.85\columnwidth]{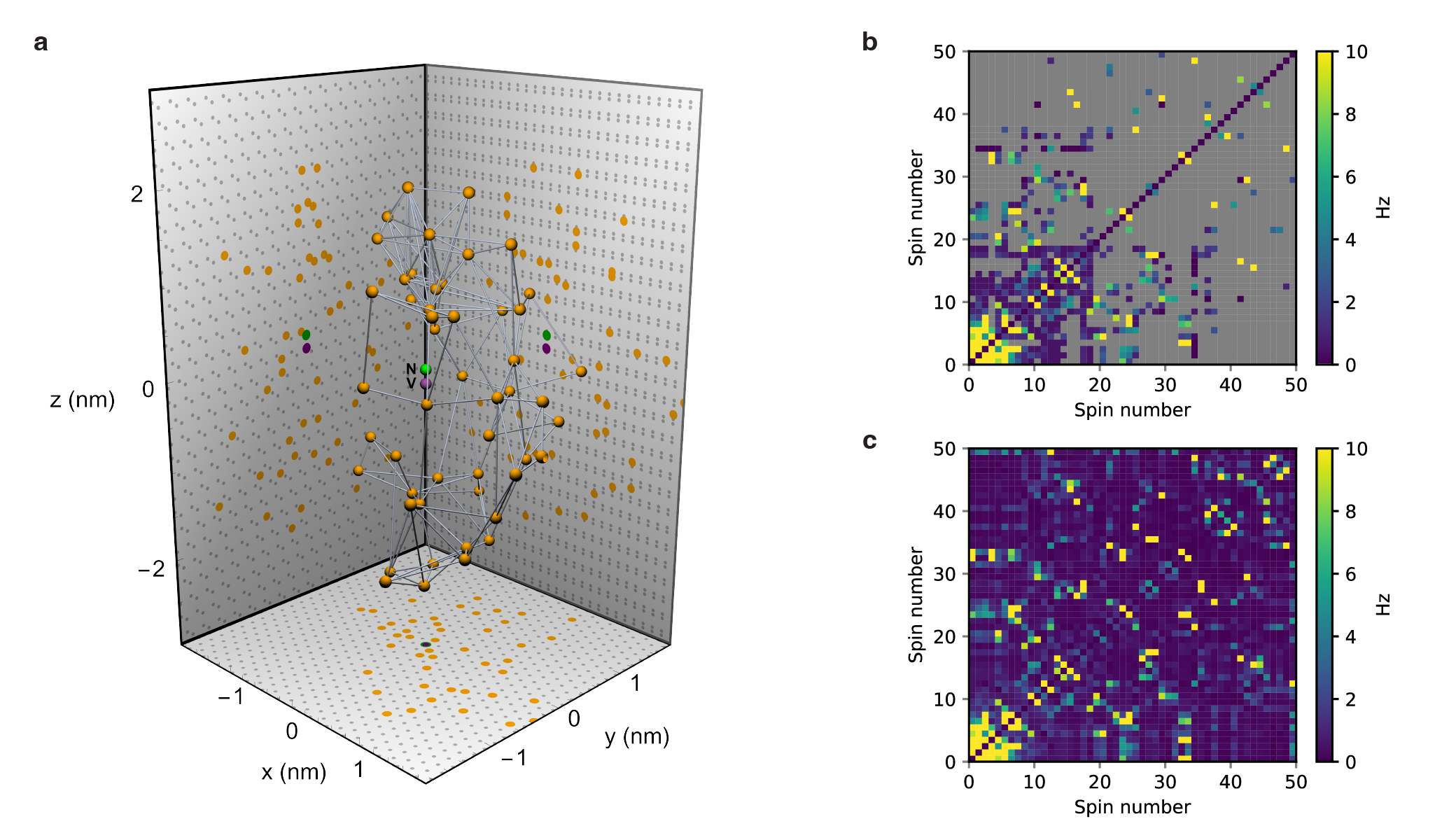}
	\end{center}
\caption{\emph{Spatial structure of the 50-spin network.} a) Most likely positions for the 50 $^{13}$C nuclear spins mapped in this work. Couplings larger than 3 Hz are visualised by the grey connections. The NV vacancy site is placed at the origin. b) Measured coupling matrix. Elements that were not measured, or did not return a clear signal (due to spectral crowding) are colored grey. c) Predicted coupling matrix, based on the most likely spin positions (Methods).}
	\label{fig_ext:positions}
\end{figure}

\begin{figure}
	\begin{center}
		\includegraphics[width=0.55\columnwidth]{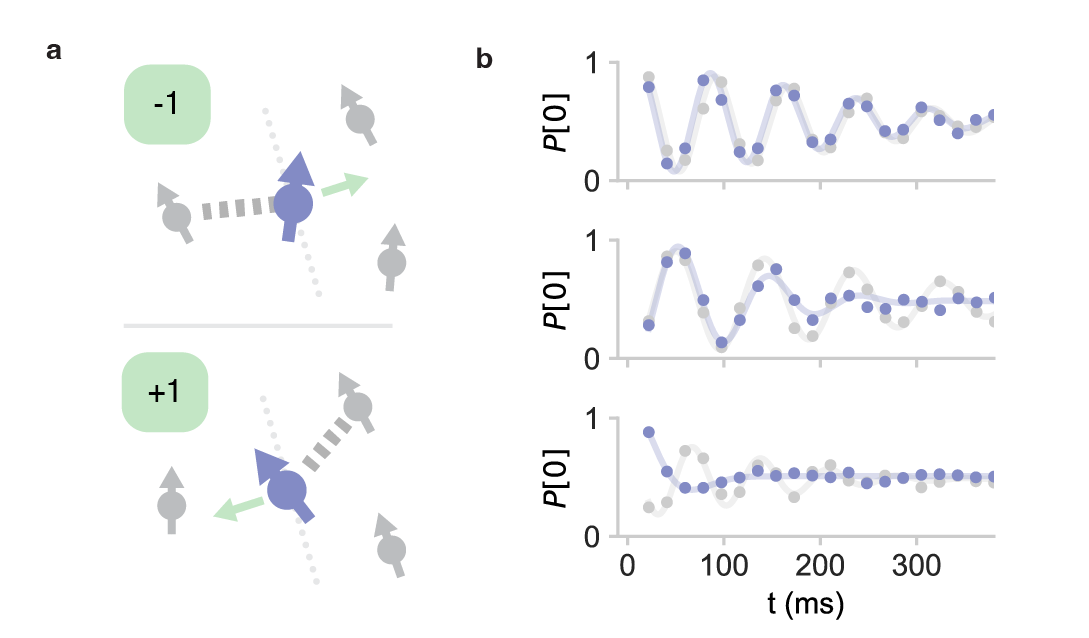}
	\end{center}
\caption{\emph{Electron-state-dependent dephasing} a) Schematic of the dephasing effect for a target nuclear spin (purple), coupled to a local spin bath (grey) under the electron-nuclear double resonance sequence (see Fig. 3). Due to the perpendicular hyperfine component (mint-green arrow), the nuclear quantisation axes change when the electron changes spin state (from $\ket{-1}$ to $\ket{+1}$). As a result, the nuclear spin couples differently to its environment in the first and second half of the spin-echo, limiting the effectiveness of the echo if the environment is in an unpolarised state. b) Experimental data of the electron-nuclear double resonance experiment (as in Fig. 3) for three different nuclear spins (C19, C18 and C5). The surrounding spin-bath is either initialised in a mixed state (purple) or polarised `up' (grey). Solid lines are fits to the data (as in Fig. 3). For some spins (most notably C5), the described dephasing effect leads to a quick drop in coherence, which can be partially regained by polarising the bath. This effect can be accurately modelled based on the extracted spin positions (see Supplementary Fig. \ref{fig:Electron_state_dependent}). The data is corrected for the difference in global and selective polarisation direction.}
    \label{fig_ext:Electron_state_dependent}
\end{figure}

\clearpage

\subsection*{Supplementary Note 1: NV system} \label{sec:sup_NV_system}
\subsubsection{Hamiltonian}
We consider the Hamiltonian of the ground-state NV electron spin, surrounded by $N$ $^{13}$C nuclei \cite{randall_manybody_2021}:
\begin{equation}
\hat{H} = \Delta_{\mathrm{ZFS}} \hat{S_z}^2 + \gee \Bz \hat{S_z}
+  \sum^{N}_{i=1} \gc \Bz  \Iz{i} 
+ \sum^{N}_{i=1} \hat{\vec{S}}  \cdot \boldsymbol{A}^{(i)} \cdot \hat{\vec{I}}^{(i)}
+  \sum^{N}_{i=1} \sum^{N}_{j = i+1} \hat{\vec{I}}^{(i)}  \cdot \boldsymbol{C}^{(ij)} \cdot \hat{\vec{I}}^{(j)} \, ,
\label{eq:full_hamiltonian}
\end{equation}
with $\Delta_{\mathrm{ZFS}} $ the zero-field splitting, $\gee$ and $\gc$ the electron and $^{13}$C nuclear gyromagnetic ratio and $\Bz$ an external magnetic field applied along the NV-symmetry axis (z-axis). Here, $\hat{\vec{S}} = (\hat{S_x}, \hat{S_y},\hat{S_z} ) $ and $\hat{\vec{I}}^{(i)} = (\Ix{i},\Iy{i},\Iz{i})$ are the electronic and nuclear spin vectors, respectively, consisting of spin-1 matrices $\hat{S_\alpha}$ and spin-$\tfrac{1}{2}$ matrices $\Ia{i} = \sa{i}/2$ (with $\sa{i}$ the Pauli spin matrices). Furthermore, $\boldsymbol{A}^{(i)}$ is the electron-nuclear hyperfine tensor and $\boldsymbol{C}^{(ij)}$ is the nuclear-nuclear dipole-dipole coupling.

For the sensing schemes presented in this work, the NV-electron spin is either in the $m_s = +1$ or $m_s = -1$ eigenstate during evolution of the nuclear spins, except for sub-$\mu$s timescales between electron pulses. As a result of the disorder induced by the hyperfine interaction $|{\Azz}^{(i)} - {\Azz}^{(j)}|\gg {C_{zz}}^{(ij)}$, nuclear flip-flops are suppressed (frozen core). Note that this condition breaks down in general when there is a high degree of spectral crowding in the system. However, in this work, we focus on spectral regions where ${C_{zz}}^{(ij)}$ (denoted as $C_{ij}$ in the main text and from here on) is generally small if $|{\Azz}^{(i)} - {\Azz}^{(j)}|$ is small, so that the condition still holds. Furthermore, based on the large zero-field splitting $\Delta_{ZFS}$ we apply the secular approximation, so that the hyperfine tensor simplifies to just the parallel ($A_{\parallel} = \Azz$) and perpendicular($A_{\perp}= \sqrt{\Azx^2 + A_{zy}^2}$) components. Considering only the nuclear Hamiltonian while the electron is in the $\pm 1$ eigenstate:

\begin{equation}
	\hat{H}_{\pm 1} = \sum^{N}_{i=1} 
    \left[ 
    ( \gc \Bz \pm A^{(i)}_{\parallel} ) \Iz{i}
    +
    A^{(i)}_{\perp} 
    ( \cos{\phip{i}} \, \Ix{i}
    + \sin{\phip{i}} \, \Iy{i})
    \right]
+  \sum^{N}_{i=1} \sum^{N}_{j = i+1} \hat{\vec{I}}^{(i)}  \cdot \boldsymbol{C}^{(ij)} \cdot \hat{\vec{I}}^{(j)} \, ,\label{eq:full_nuclear_hamiltonian}
\end{equation}
where $\phip{i}$is the perpendicular hyperfine azimuthal angle. Here, we neglect the small correction on $C_{ij}$ due to the electron spin state ($C^{+}_{ij} = C^{-}_{ij}$) \cite{abobeih_atomicscale_2019}. Supplementary Note 4 discusses the non-neglible effect of this correction observed in the specific experiments.

Under the application of a strong magnetic field ($\gc \Bz \pm A_{\parallel} \gg A_{\perp}$), we can further simplify the Hamiltonian, with the purpose of generalizing the effective dynamics of our system. To this end, we include the perpendicular hyperfine component ($A_{\perp}$) as a correction to the $\Iz{i}$ terms, describing the increased nuclear precession frequency \cite{taminiau_detection_2012}. This results in Eq. 1 in the main text:
\begin{equation}
\hat{H}_{\pm1} \approx \sum^{N}_{i=1} A^{\pm}_{i} \Iz{i} 
+ \sum^{N}_{i=1} \sum^{N}_{j = i+1} C_{ij} \Iz{i} \,\Iz{j} \,,
\label{eq:eff_nuclear_hamiltonian}
\end{equation}
with $A^{\pm}_{i} = \sqrt{ (\gc \Bz \pm A^{(i)}_{\parallel})^2 + (A^{(i)}_{\perp})^2  } $ the effective nuclear spin frequencies. In the main text, $A_i$ describes the spin frequencies when the electron is in the $m_s  = -1 $ state ($\pm$-sign is omitted for simplicity).

\subsubsection{Initialisation}
For the experiments in the main text, the nuclear spins are polarised via two techniques. We employ a combination of dynamical nuclear polarisation (PulsePol, \cite{schwartz_robust_2018}) and SWAP sequences with the electron spin \cite{bradley_tenqubit_2019}. The latter method typically yields a higher degree of polarisation \cite{bradley_tenqubit_2019}, but is restricted to a limited number of spectrally isolated nuclear spins (see Supplementary Table \ref{tab:spin_overview}). If any of these SWAP-initialised spins participate in a sensing chain, we re-initialise them right before the final SEDOR-yx sequence to maximize the signal. All other spins are polarised via the PulsePol sequence, which is known to produce varying degrees of polarisation \cite{randall_manybody_2021,rao_spinbath_2020,villazon_integrability_2020}. Following the definitions given in the supplement of Ref. \cite{randall_manybody_2021}, we set $\tau = \SI{0.434}{\micro \second}$, resonant with the nuclear Larmor frequency, $N = 4$ and choose $R$ in the range $500 - 10000$, dependent on the rate of polarisation of the spins in the chain. For resonant reset of the electron state after each PulsePol step, the repump laser power was set to \SI{1000}{\nano \watt}, with a repump time of  \SI{5}{\micro \second}. For some experiments (in particular the 2D spectroscopy) the laser power was reduced to $\SI{333}{\nano \watt}$ (and the repump time increased to \SI{10}{\micro \second}), to limit electron ionisation for $R > 5000$. 

For the analytical expressions derived in the supplement, we capture the varying degree of polarisation of $N$ nuclear spins by a partially mixed system initial state:

\begin{equation}\label{eq:DNP_initial_state}
    \hat{\rho_0}=\hat{\rho}_{e,0}\bigotimes_{i=1}^N \frac{1}{2} \left(\ide + p_i \Iz{i} \right),
\end{equation}
with $\ide$ the (two-dimensional) identity matrix, the polarisation degree $p_i \in [-1,1]$ for spin $i$ and $\hat{\rho}_{e,0}$ the initial state of the NV electron spin. The $p_i$ can be obtained by independent state preparation and measurement characterisations \cite{randall_manybody_2021}.

\subsubsection{Readout}
For all experiments presented in this work, the signal is read out via dynamical-decoupling sensing sequences (DD or DDRF \cite{taminiau_detection_2012,bradley_tenqubit_2019}) with the electron spin. Only a limited number of spins can be read out selectively (see Supplementary Table \ref{tab:spin_overview}), due to spectral overlap between nuclear spins. Hence, we choose the first spin in all sensing chains to be one that is directly accessible to the electron spin. Combined initialisation and readout fidelity varies between spins ($0.44(2) - 0.95(2)$, corrected for electron readout fidelity).

\clearpage
\subsection*{Supplementary Note 2: Spectral crowding in the NV-nuclear system}
\setcounter{subsubsection}{0}

\label{sec:spectral_crowding}
The schematic in the main text (Fig. 1), describes the challenge of mapping a spectrally crowded spin network in a general sense, for an abstract spin network described by frequencies $A_i, C_{ij}$ (Eq. 1). In the system studied here, all nuclear spins are $^{13}$C spins and the nuclear spin frequencies $A_i$ are set by the coupling to the NV electron spin. Therefore, there is a specific relationship between the nuclear-spin frequencies and their 3D position with respect to the electron spin (see also Supplementary Fig. \ref{fig_ext:spectral_regions}). In this section we discuss how different spectral regions can be defined, and which regions can and cannot be accessed with different methods.

To understand which parts of the network can be mapped with the new methods, and where different methods break down, we define three spectral regions: \emph{isolated}, \emph{spectrally crowded} and \emph{spatially crowded}. The \emph{isolated} region is the set of spins $i$, for which:
\begin{equation} \label{eq:isolated_region}
    |A_i - A_j| >   1/{\tstar} \,\, \forall\,\, j \,,
\end{equation}
which states that the frequency of spin $i$ is unique. Here, we define this condition as being satisfied if there are at least 4 spectral widths (s.d.) between the resonances of two spins. In a natural abundance (1.1\%) sample, we calculate that 20(3) spins typically satisfy this condition, and only about 8(3) are isolated by more than $\SI{2}{\kilo \hertz}$ from any other spin (assuming $\tstar \sim 5$ ms and a purely dipolar hyperfine interaction). In the NV-nuclear system, only this set of spins can be read out selectively with the electron spin \cite{bradley_tenqubit_2019, abobeih_atomicscale_2019, taminiau_detection_2012} (ignoring notable exceptions in the form of strongly-coupled spin-pairs \cite{abobeih_onesecond_2018,bartling_entanglement_2022}), even though the electron spin typically couples to the majority of nearby nuclear spins ($\deltah{i} > 1/T_{2,e}$). 

Next, we define the \emph{spectrally crowded} region as the set of spins $i$ for which:
\begin{equation} \label{eq:spectrally_crowded_region}
    |A_i - A_j| < 1/{\tstar} \implies C_{ij} \lesssim 1/{\ttwo} \,\, \forall\,\, j \,,
\end{equation}
meaning that spins may overlap spectrally, but if they do, they are typically not coupled strongly together. In the NV-nuclear system, this region includes nuclear spins which have similar hyperfine interaction with the electron, but are \emph{spatially} separated, for example when they are on opposite sides of the NV center in the $x,y$-plane. 

Finally, the \emph{spatially crowded} region includes all other spins, which may be overlapping spectrally as well as coupling strongly together. 

We schematically draw the three spectral regions in Supplementary Fig. \ref{fig:spectral_regions}, with the bottom color bar denoting the three spectral regions (green $=$ \emph{isolated}, orange $=$ \emph{spectrally crowded} and red $=$ \emph{spatially crowded}). The blue through violet bands indicate spin frequencies $A_i$ (analogous to the colored circles in the main text), shifted by the hyperfine interaction to the electron spin. In the schematic, we simplify the more complex dipolar isoplane shape \cite{perunicic_quantum_2016} and consider only the radial dependence ($A_i \propto r^{-3}_i$). A particular nuclear-spin-network configuration is drawn as an example. 

Only limited information can be attained from the network using pairwise SEDOR sequences \cite{abobeih_atomicscale_2019}. If we allow no assumptions on the underlying coupling structure, pairwise measurements can only unambiguously assign couplings to the spins in the isolated region (coloured white). Using SEDOR, it is possible to measure couplings between one spin in the \emph{isolated} region and one in the \emph{spectrally crowded} region. However, those couplings cannot be assigned to a spin in the latter region without resorting to a detailed microscopic model \cite{abobeih_atomicscale_2019}. As an example, using SEDOR, we find that the frequencies $A_1, A_2$ and $A_3$ all exhibit coupling to some spin at $A_4$ (coloured grey), but the couplings could belong to either of the spins at $A_4$. Also, couplings for which both spins lie in the \emph{spectrally crowded} region (coloured black) cannot be accessed, as the spin-selective readout with the electron spin breaks down in this region.

\begin{figure}[ht]
	\begin{center}
		\includegraphics[width=1\columnwidth]{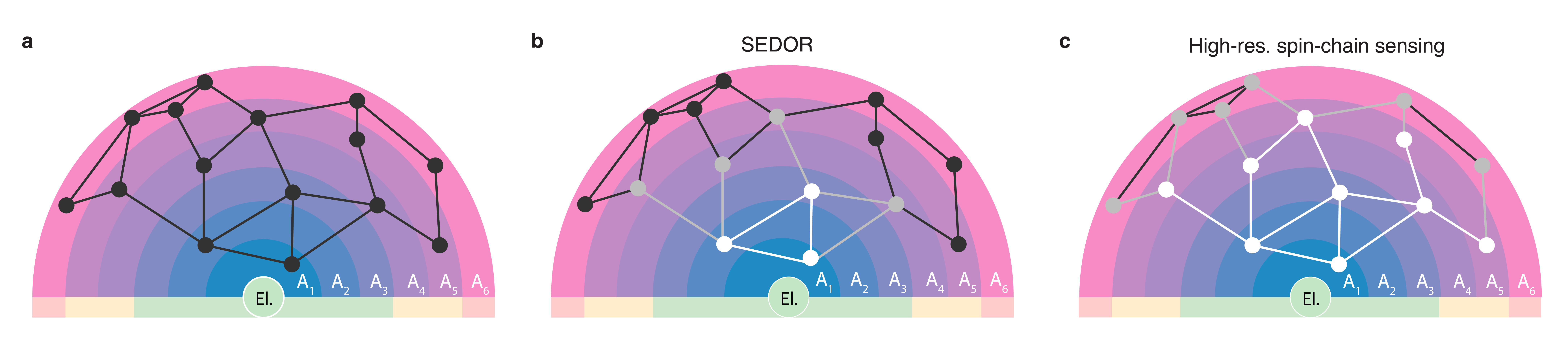}
	\end{center}
\caption{\textbf{Unlocking spectral regions.} \textbf{a} Alternative representation of Fig. 1, specific to the electron-nuclear system. Nuclear spins are drawn as black dots, connected by lines denoting observable couplings ($C_{ij} > 1/{\ttwo}$). The colored bands indicate spin frequencies $A_i$, shifted by the hyperfine interaction ($\deltah{i}$) with the NV electron spin (mint green). The green, orange and red color scales at the bottom denote three spectral regions: \emph{isolated}, \emph{spectrally crowded} and \emph{spatially crowded}, respectively. In the \emph{isolated} region, spin frequencies are well-separated in frequency, (single spin per band) and can be read out directly with the electron \cite{bradley_tenqubit_2019}. In the \emph{spectrally crowded} region (orange), multiple spins may occupy a single frequency band, but do not couple among each other. In the \emph{spatially crowded} region (red), spins may also couple to spins within their frequency band, resulting in a decreased spin-echo coherence. \textbf{b} Spins that can be mapped using standard pairwise SEDOR \cite{abobeih_atomicscale_2019}. The color of the coupling line indicates whether it can be measured and conclusively assigned (white), measured but not assigned (grey), or not measured at all (black). \textbf{c} Spin-chain sensing increases the number of couplings that can be measured and assigned (e.g. when \emph{both} spins are in the \emph{spectrally crowded} region) and unlocks spins in the \emph{spatially crowded} region. Additionally, the high-resolution measurements of $\deltah{i}$ allow for assigning otherwise ambiguous couplings in the \emph{spectrally crowded} region.}
    \label{fig:spectral_regions}
\end{figure}
\subsubsection{Spectral regions}\label{seq:spectral_regions}

The spin-chain sensing (Fig. 2) unlocks new parts of the crowded region that can be mapped (Supplementary Fig. \ref{fig:spectral_regions}c). Consider the previously discussed couplings between $A_1, A_2, A_3$ and $A_4$. If we measure a chain connecting $A_1, A_4$ and $A_2$, we conclude that $A_1$ and $A_2$ couple to the same spin at $A_4$. Measuring a chain between $A_1, A_4$ and $A_2$ does not result in an observable coupling, so we conclude that $A_2$ is coupled to another spin at $A_4$. Note that this reasoning relies on the fact that it is possible to assert that $A_1, A_2$ and $A_3$ couple only to a \emph{single} spin at $A_4$. Experimentally, this can be verified by observing a single dominant oscillation in the signal (instead of beatings or decay).

Besides assigning measured couplings to spins, spin-chain sensing allows for access to an extended number of couplings, particularly in the \emph{spectrally crowded} and \emph{spatially crowded} regions (Supplementary Fig. \ref{fig:spectral_regions}c). For example, by measuring a looped spin chain through frequencies $A_1, A_2, A_4, A_5, A_3, A_1$, we access the coupling between two spins in the spectrally crowded region (at $A_4$ and $A_5$) and directly map the connectivity of the 5 spins in the loop. In addition, we can use the newly unlocked spins (at $A_4$ and $A_5$) to probe couplings to the \emph{spatially crowded} region (grey spins at $A_6$). Finally, by increasing the spectral resolution (Figs. 3 and 4), we resolve remaining ambiguity in the \emph{spectrally crowded} region. 

Even though the high-resolution spin-chain sensing fully unlocks the \emph{spectrally crowded} region and allows us to probe the \emph{spatially crowded} region, the latter also imposes a limit on the applicability of the method. In particular, the spin-chain sensing relies on an extended nuclear spin-echo coherence time (ideally from $\tstar$ to $\ttwo$). However, in this region, a decoupling pulse inadvertently also acts on other nuclei, so that their spin-spin couplings are retained. This results in the re-emergence of quasi-static ($\tstar$-like) noise also known as instantaneous diffusion \cite{tyryshkin_electron_2012}, limiting the spin-echo coherence time $T_{2,\mathrm{SE}}$ to:
\begin{equation}
    \tstar \leq T_{2,\mathrm{SE}} \leq \ttwo \, .
\end{equation}
In the case when many strongly coupled spins reside in the same frequency band (see Supplementary Eq. \ref{eq:spectrally_crowded_region}), we expect the coherence to be effectively reduced to $\tstar$, rendering the effect of the double resonance sequence useless. Hence, utilizing spins inside the \emph{spatially crowded} region as probes of their environment is infeasible (except for strongly interacting spins, if $C_{ij}>1/\tstar$), which sets the limit of the functional range of spin-chain sensing.

\subsubsection{Numerical simulations}
\begin{figure}[t]
	\begin{center}
		\includegraphics[width=0.55\columnwidth]{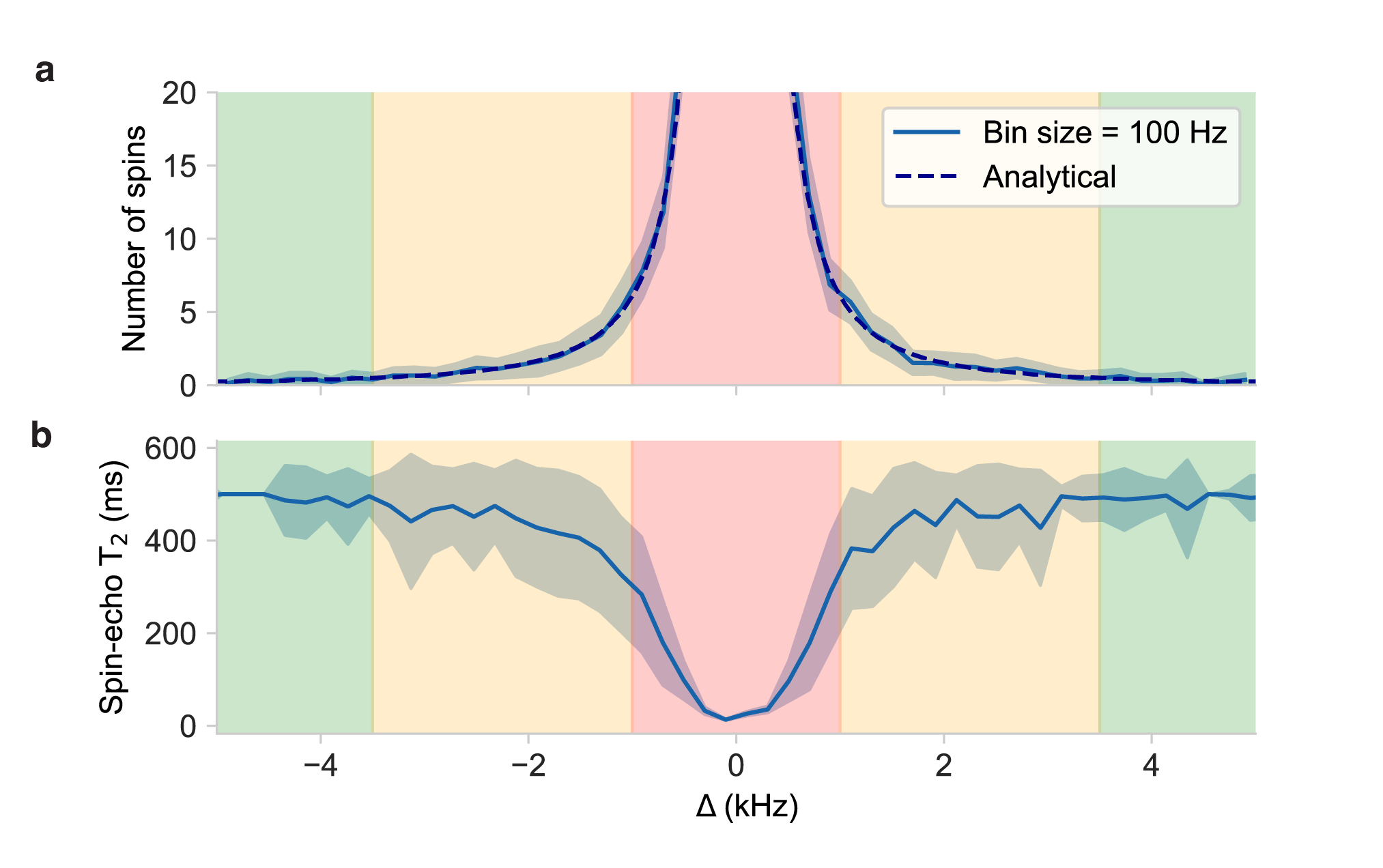}
	\end{center}
\caption{\textbf{Spectral spin-density and spin-echo coherence reduction.} \textbf{a} Expected spectral density (number of spins per 100 Hz) as a function of the hyperfine shift $\deltah{}$, assuming dipolar electron-nuclear coupling. The blue solid line denotes the mean of 100 randomly generated systems (containing $>15000$ spins). The analytical expression in Supplementary Eq. \ref{eq:spin_density_analytical} (dashed line) describes the simulated distribution well. \textbf{b} Expected spin echo coherence time (Eff. $\ttwo$) as a function of the hyperfine shift $\deltah{}$ (considering systems of $>1000$ spins), taking into account only quasi-static effects and limiting the maximal coherence to $\ttwo = 500$ ms. Coherence is reduced as the pulses in the spin-echo are resonant with multiple spins (see (a)), resulting in a recoupling of nearby spins. Here, the Rabi frequency is set to $f_{\mathrm{Rabi}} \approx 1/{2 \tstar}$. Broader pulses (higher $f_{\mathrm{Rabi}}$) increase the number of recoupled spins, further reducing the nuclear coherence. Shaded areas denote the spread (one s.d.) between simulated systems .}
    \label{fig:spin_density}
\end{figure}

To quantitatively investigate the regions visualised in Supplementary Fig. \ref{fig:spectral_regions}, we perform Monte Carlo simulations of randomly generated NV-nuclear systems. First, we compute the spectral spin density (i.e. number of spins within a frequency bin) as a function of the hyperfine shift $\deltah{}$ (Supplementary Fig. \ref{fig:spin_density}a). We find that the mean spin density is well described by (dotted line):
\begin{equation}\label{eq:spin_density_analytical}
    \bar{\rho}(\deltah{}) =   \frac{\pi^2 \, \alpha \, \rho_0}{{\deltah{}}^2}  \, ,
\end{equation}
with $\bar{\rho}$ the mean spin density in frequency space (Hz$^{-1}$), $\rho_{0} = 1.950 \, \mathrm{nm}^{-3}$ the spatial $^{13}$C density and $\alpha = \mu_0 \gee \gc \hbar / 4\pi$, with $\mu_0$ the Bohr magneton, $\hbar$ the reduced Planck constant and $\gee$ and $\gc$ the electron and carbon nuclear gyromagnetic ratios, respectively. Supplementary Fig. \ref{fig:spin_density}a shows the expected number of $^{13}$C spins within a frequency bin of $\SI{100}{\hertz} \sim 1/(2 \, \tstar)$, which gives a measure for the probability to find spectrally overlapping spins. As described above, here we choose to define the start of the \emph{spectrally crowded} region at the condition:
\begin{equation}
    \bar{\rho} \approx \tstar \, , %\frac{0.5}{2 \, \tstar } \, ,
\end{equation}
yielding $|\deltah{}| \approx  \SI{3.5}{\kilo \hertz}$. For larger $|\deltah{}|$, we expect on average less than one spin per $\tstar$-limited frequency bin (\emph{isolated region}, see green region in Supplementary Fig \ref{fig:spin_density}a).

To quantify the transition between the \emph{spectrally crowded} and \emph{spatially crowded} regions, we compute the expected drop in spin-echo coherence, taking into account instantaneous diffusion \cite{tyryshkin_electron_2012} (Supplementary Fig \ref{fig:spin_density}b). Assuming linear addition of the dephasing rates \cite{bauch_ultralong_2018}:
\begin{equation}
    1/T_{2,\mathrm{SE}} = 1/\ttwo + 1/T_{2,\mathrm{ID}} \, ,
\end{equation}
with $1/T_{2,\mathrm{ID}}$ the instantaneous diffusion dephasing rate and $\ttwo \approx \SI{500}{\milli \second}$ the isolated spin-echo coherence time  \cite{bradley_tenqubit_2019}. We compute $T_{2,\mathrm{ID}}$ by examining the mean $\tstar$ of the subsystem of spins that occupy a frequency bin of size 100 Hz for all $\deltah{}$. We mark the start of the \emph{spatially crowded} region at the point $T_{2,\mathrm{SE}} \approx 0.5 \,\ttwo$, which implies $|\deltah{}| \lesssim \SI{1}{\kilo \hertz}$ (Supplementary Fig. \ref{fig:spin_density}b). For the 23 newly characterised spins in this work, most belong to the \emph{spectrally crowded} or \emph{spatially crowded} regions (8 satisfy $3.5 < |\deltah{i}| < 7.5$ kHz, 10 satisfy $1 < |\deltah{i}| < 3.5$ kHz, and 4 satisfy $|\deltah{i}|  < 1$ kHz, see Supplementary Table \ref{tab:spin_overview}).

The discussion so far has focused on the degree of spectral crowding using the numerical values for a natural abundance  sample ($1.1 \%$ $^{13}$C). We now examine how these relations depend on the isotope concentration. In the basic picture of a point-dipole electron spin, surrounded by a bath of nuclear spins with spatial density $\rho_0$, the system exhibits a form of scale invariance. That is, we can define the (dimensionless) spectral density: 
\begin{equation}\label{eq:independent_spectral_density}
    \frac{\bar{\rho}(\deltah{})}{\tstar} = \frac{\pi^2 \, \alpha \, \rho_0}{\tstar {\deltah{}}^2}  \, ,
\end{equation}
describing the number of spins per line width. For dipolar interactions, both the electron-nuclear hyperfine shift $\deltah{}$ as well as the nuclear line width ($\sim 1/\tstar$) scale linearly with the density, so that the dimensionless spectral density (Supplementary Eq. \ref{eq:independent_spectral_density}) is independent  of $\rho_0$. Intuitively, this can by understood as both the nuclear-spin line widths and the spacing between nuclear-spin frequencies scaling with $\rho_0$, keeping their ratio (i.e. the degree of spectral crowding) constant. In principle, in this elementary system, the physics and quantities like the number of spins that can be mapped are independent of concentration, with only absolute time and distances being rescaled.

In practice, however, this scale invariance breaks down for both for low and high isotope concentrations. At low isotope concentrations, other noise sources will start to limit both the nuclear line widths (via reduced $\tstar$), as well as the spin-echo coherence times (via reduced $\ttwo$). At high concentrations --- including around the natural $1.1\%$ abundance --- the discreteness of the lattice and the contact hyperfine interaction due to the finite NV electron wave function need to be taken into account. Predictions for the optimal concentration for a given goal, such as controlling the largest network, likely need detailed numerical simulations taking these details into account, which we do not pursue here.

\clearpage
\subsection*{Supplementary Note 3: Spin-chain sensing} \label{sec:sup_spin_chain_signal}
\setcounter{subsubsection}{0}

\subsubsection{Signal analysis}
 Here, we analyse the system evolution under the spin-chain sensing sequences developed in this work, and give analytic expressions for the expected resulting signals. Given the initial state in Supplementary Eq. \ref{eq:DNP_initial_state}, we calculate the $z$-expectation value of the first nuclear spin ($1$) in the chain, for a chain of length $N$ after applying the concatenated SEDOR sequence (Fig. 2c). We analyse the evolution of the system by dividing it up into separate blocks, set by the subsequent SEDOR sequences. As the nuclear spins are initialized along the $z$-axis (have no off-diagonal component, see Supplementary Eq. \ref{eq:DNP_initial_state}), the spins in the chain do not evolve, except for those participating in a SEDOR block. Therefore, we can restrict our analysis to the subspace spanned by the spins in each block.
 
 We find a recursive expression for the evolution of two subsequent spins ($j+1$ and $j$) under a SEDOR block (for both `xx' and `yx' type sequences). To this end, we trace out the $j+1$-subspace, as only the density matrix of spin $j$ is needed to calculate the subsequent SEDOR evolution between spin $j$ and $j-1$. This allows us to find two recursive formulas involving only the diagonal elements of each spin density matrix. By applying these expressions $N-1$ times, we retrieve the $z$-expectation value of spin $1$, which is a function of all nuclear-nuclear couplings in the chain.
 
Evolution during a single SEDOR-xx or SEDOR-yx block for two subsequent spins in the chain can be described by the unitary:

\begin{equation}
\Uxk{j+1}{j} = \pitwox{j} ~ \Um ~ \pix{j+1} ~ \pix{j}  ~ \Um ~ \pitwok{j} ,
\end{equation} \label{eq: Usedor-xj}

with $\Rk{j}$ a rotation of spin $j$ by an angle $\theta$ around axis $k \in \{x,y\}$ and $\Um$ the free evolution under the Hamiltonian $\hat{H}_{-1}$ in Supplementary Eq. \ref{eq:eff_nuclear_hamiltonian}, considering only spin $j+1$ and $j$ \cite{randall_manybody_2021}. Without loss of generality, we can write the initial state of any two subsequent spins as:

\begin{equation}
\rho_{j+1,j}(0) = \frac{1}{4}
   \begin{pmatrix}
    1 + \alpha_{j+1} & \beta_{j+1} \\
    \beta^{*}_{j+1} & 1 - \alpha_{j+1}  \\
    \end{pmatrix}
\otimes 
   \begin{pmatrix}
    1 + p_j & 0 \\
    0 & 1 - p_j \\
    \end{pmatrix}
\,,
\end{equation}
where the first density matrix denotes the subspace of spin $j+1$, which can be in any arbitrary quantum state and the second density matrix describes the subspace of spin $j$, initialised according to Supplementary Eq. \ref{eq:DNP_initial_state}. We let the system evolve for time $t$ under $\Uxk{j+1}{j}$, after which we trace out the $j+1$ subspace, resulting in the density matrix for spin $j$:
\begin{align*}
    \rho_{j}(t) & = \Tr_{j+1}\left(\Uxk{j+1}{j} ~ \rho_{j+1,j}(0) ~  {\Uxk{j+1}{j}}^{\dagger} \right) \\
    & = \frac{1}{2}\begin{pmatrix}
1 + \aj{k}{} & \bj{k}{} \\
\bjs{}{k} & 1 - \aj{k}{}  \\
\end{pmatrix} \,,
\end{align*}
leading to the following update rule for the diagonal density matrix elements of spin $j$ under SEDOR-xx and SEDOR-yx:
\begin{align}
    \aj{}{x} &= p_j \cos{\frac{2 \pi C t}{2}} \,, \label{eq:ajx} \\ 
    \aj{}{y} &= p_j \sin{\frac{2 \pi C t}{2}}\, \alpha_{j+1} \label{eq:ajy} \,, 
\end{align}
with $C = C_{j,j+1}$ the coupling between the spins in $\si{\hertz}$. Note that the off-diagonal terms $\beta_{j+1}$ drop out when we only consider the diagonal elements of spin $j$. To calculate the $z$-expectation value of the first spin in the chain after a SEDOR-xx block and $N-2$ concatenated SEDOR-yx blocks according to Fig. 2c, we iteratively apply Supplementary Eqs. \ref{eq:ajx} and \ref{eq:ajy} to find:
\begin{equation} \label{eq:spin_chain_signal_ideal}
    \expval{\Iz{1}} = \frac{1}{2} \, \alpha_{1} = 
    \frac{1}{2} \, p_{N-1}  \cos{\left(\pi C_{N-1,N} t_{N-1,N} \right)}
    \prod_{j=1}^{N-2} p_j \sin{\left( \pi C_{j,j+1} t_{j,j+1}\right)} \, ,
\end{equation}
\subsubsection{Decoherence of the chain} \label{sec:sup_chain_decoherence}
Supplementary Eq. \ref{eq:spin_chain_signal_ideal} does not take into account any imperfections due to decoherence or pulse errors. In the following, we model the effect of decoherence, which is the main factor limiting the signal. Here, we do not take into account the effect of pulse errors, but this can be implemented analogous to Ref. \cite{randall_manybody_2021}.

We model decoherence by multiplying the signal of each SEDOR block by an exponential decay function, parameterised by a characteristic spin-echo decay time $\tau_j$. In the case that the spin-echo in the SEDOR is perfectly effective, meaning that spin $j$ is fully decoupled from all other nuclei, the decay is governed by dynamic noise sources ($\tau_j  \sim 250 - 800 $ ms) \cite{bradley_tenqubit_2019}. However, for a spin at a spectrally crowded frequency, the decoupling pulse inadvertently also acts on other nuclei, so that their coupling to spin $j$ is retained. This results in the re-emergence of quasi-static noise (instantaneous diffusion) discussed in Supplementary Note 2, which we model by adding a Gaussian decay to Supplementary Eq. \ref{eq:spin_chain_signal_ideal}:

\begin{equation} \label{eq:spin_chain_signal_decay}
    \expval{\Iz{1}} = 
    \frac{1}{2} \, p_{N-1}  \cos{\left(\pi C_{N-1,N} t_{N-1,N} \right)} \,
    e^{-\left(\frac{t_{N-1,N}}{\tau_{N-1}}\right)^2}
    \prod_{j=1}^{N-2} p_j \sin{\left( \pi C_{j,j+1} t_{j,j+1} \right)} \,
    e^{-\left(\frac{t_{j+1,j}}{\tau_{j}}\right)^2}
    \,,
\end{equation}
Even though the recoupled spins are partially polarised, we expect them to impart only a decay and no frequency shift on the signal, as they will also undergo the $\tfrac{\pi}{2}$-pulse, negating any $z$-axis polarisation. For the experiments in Fig. 2d-f, we set $t_{i,i+1} = \tfrac{1}{2} C^{-1}_{i,i+1}$ for all SEDOR-yx blocks, which reduces Supplementary Eq. \ref{eq:spin_chain_signal_decay} to:
\begin{equation} \label{eq:decay_fit}
    \expval{\Iz{1}} = A_{N-1} \cos{\left(\pi C_{N-1,N} t_{N-1,N} \right)} \,
    e^{-\left(\frac{t_{N-1,N}}{\tau_{N-1}}\right)^2}
    \, ,
\end{equation}
with $A_{N-1} = \frac{1}{2} p_{N-1} \prod_{j=1}^{N-2} p_j e^{-\left(\frac{C_{j+1,j}}{2 \tau_{j}}\right)^2}$ the signal amplitude. We use Supplementary Eq. \ref{eq:decay_fit} to fit the data in Fig. 2d-f with free parameters $A_{N-1},C_{N-1,N}, \tau_{N-1}$ and an arbitrary offset. Note that even though the signal amplitude $A_{N-1}$ is affected by the coherence and polarisation of all spins in the chain, the spectral resolution with which $C_{N-1,N}$ can be determined is only limited by the coherence of spin $N-1$ (i.e. $\tau_{N-1}$). The decay of the signal $A_{N-1}$ with increasing number of spins due to the imperfect polarisation ($p_j<1$) and finite coherence times $\tau_{j}$ limits how long a chain can be practically formed. Note that for spins further from the NV $\tau_{j}$ tends to decrease due to imperfect decoupling (instantaneous diffusion in the spatially crowded region, see Supplementary Note 2), ultimately limiting the range for high-resolution sensing. 

\clearpage
\subsection*{Supplementary Note 4: Electron-nuclear double resonance sequence}
\setcounter{subsubsection}{0}

\label{sec:sup_ENES_analysis}
\subsubsection{Signal analysis}
Next, we analyse the system evolution under the electron-nuclear double resonance sequence (see Fig. 3). To this end, we consider the interaction between the electron (`el') and nuclear spin $j$ (Fig. 3) by considering the following unitary: 
\begin{equation} \label{eq:U_ENES}
\Uxx{\mathrm{el}}{j} =  \pitwox{j} ~ \Up ~ \pix{j}  ~ \Um ~ \pitwox{j} \,,
\end{equation}
with $U_{\pm}$ the unitary evolution under Hamiltonian $\hat{H}_{\pm 1}$ (Supplementary Eq. \ref{eq:eff_nuclear_hamiltonian}). We compute the $z$-expectation value of nuclear spin $j$ after applying $\Uxx{\mathrm{el}}{j}$ (starting in initial state given by Supplementary Eq. \ref{eq:DNP_initial_state}):
\begin{equation} \label{eq:ENES_signal}
    \expval{\Iz{j}} = \frac{1}{2} p_j \cos{\left( 2\pi \deltah{j} \, t  \right)} \,.
\end{equation}
Note that for some experiments in this work, the final $\pi/2$-rotation was performed along the $-x$-axis instead of the $x$-axis, which leads to a minus sign on the signal, but has no impact on the frequency or amplitude. The frequency (in Hz) is defined as the hyperfine shift referred to in the main text:
\begin{equation}
    \deltah{j} = \frac{1}{2} (A^{-}_{j} - A^{+}_{j}) \, ,
\end{equation}

$\deltah{j}$ provides a high-resolution measurement of the frequency shift due to the electron-nuclear hyperfine interaction. In this work, the function of this measurement is to distinguish different spins with similar precession frequencies. That is, $\deltah{j}$ provides a high-resolution label for the spins. 

An additional application is to perform precise spectroscopy of the system to determine the hyperfine interaction, for example for comparison to density functional theory calculations \cite{nizovtsev_nonflipping_2018} or to determine the Hamiltonian parameters for developing precise quantum control. Next, we analyze the relation of $\deltah{j}$ to the hyperfine parameters.

We use the spin frequencies $A^{\pm}_{j}$ introduced in Supplementary Eq. \ref{eq:eff_nuclear_hamiltonian}, but now allow for a slight misalignment of the external magnetic field away from the NV-axis (z-axis):

\begin{equation}\label{eq:ENES_non_secular}
        \deltah{} = \frac{1}{2} \left( \sqrt{ (\gc \Bz - \Azz)^2  + (\gc B_x - \Azx)^2 + (\gc B_y - A_{zy})^2  } - \sqrt{ (\gc \Bz + \Azz)^2 + (\gc B_x + \Azx)^2 + (\gc B_y + A_{zy})^2  } \right) \,,
\end{equation}
where we omit the spin-subscript $j$ for readability. Here, $B_x$ and $B_y$ are the perpendicular field components. Note that for simplicity we use a purely geometric argument and do not take into account spin mixing for the eigenstates (i.e. the eigenstates are set as the electron and nuclear spin states), which would introduce additional (small) frequency shifts. Supplementary Eq. \ref{eq:ENES_non_secular} shows that measuring $\deltah{}$ for different magnetic field vectors makes it possible to determine the hyperfine parameters, given that the field components are known. As the magnetic field components generally are not exactly known, we now analyze various situations and approximations. 

For a strong field aligned along the z-axis, the perpendicular hyperfine components are a small perturbation:
\begin{align}\label{eq:Delta_j_approx}
        \deltah{} &= \frac{1}{2} \left( \sqrt{ (\gc \Bz - \Azz)^2 + {A_\perp}^2  } - \sqrt{ (\gc \Bz + \Azz)^2 + {A_\perp}^2  } \right) \\
        &\approx - \Azz \left( 1 -  \frac{{A_\perp}^2}{2(\gc \Bz - \Azz)(\gc \Bz + \Azz)} \right) 
        \, .
\end{align}
Supplementary Eq. \ref{eq:Delta_j_approx} shows that the measurement predominantly probes the hyperfine component parallel to the magnetic field. For the magnetic field used in this work ($\Bz \sim 403$ G) and typical hyperfine values  ($\Azz \sim A_\perp \sim 10$ kHz), the typical deviation from $\Azz$, due to $A_\perp$, is smaller than $0.03$ \% (less than 3 Hz). Note that the effect of a finite $A_\perp$ is suppressed because both terms in the top line of Supplementary Eq. \ref{eq:Delta_j_approx} tend to shift in the same manner.

A misaligned field, with non-zero $B_x$ and $B_y$ components, in combination with a non-zero $A_\perp$, causes an additional frequency shift of $\deltah{}$. Due to the sign difference in the first and second terms in Supplementary Eq. \ref{eq:ENES_non_secular}, the effect is relatively large and a perpendicular field of $\sim 0.5$ G causes a frequency shifts of a few Hz. 

These results show that additional measurements and/or analysis are required to fully exploit the high-spectral-resolution measurements presented here for precision spectroscopy of the Hamiltonian parameters. However, this does not affect the capability used in this work to resolve different spins with high spectral resolution.

\subsubsection{Pulse errors}
To investigate how pulse errors affect the measured frequency $\deltah{}$, we model the first $\pi/2$-pulse and the spin-echo $\pi$-pulse on nuclear spin $j$ as imperfect X-rotations with excitation probability $f^2$:
\begin{equation}
\Rx{j}{f} = \sqrt{1-f^2} ~\ide  - i f \Ix{j} \, ,
\end{equation}
leading to the adapted unitary (Supplementary Eq. \ref{eq:U_ENES}):
\begin{equation}
\Uxx{\mathrm{el}}{j} =  \pitwox{j} ~ \Up ~ \Rx{j}{f_2}  ~ \Um ~ \Rx{j}{f_1} \,.
\end{equation}
We find the signal contains three frequency components:
\begin{equation}
 \expval{\Iz{j}} = \kappa_1 \cos{(\deltah{j} \, t)} + \kappa_2 \cos{(\omega_\mathrm{L} \, t)} + \kappa_3 \cos{\left(\frac{1}{2} A^{+}_{j} \, t\right)} \, ,   
\end{equation}
with $\kappa_1$, $\kappa_2$, $\kappa_3$ some real constants, determined by the pulse excitation probabilities $f^2_1$ and $f^2_2$. For perfect $\pi/2$-pulse and $\pi$-pulse, only the first term remains, corresponding to the hyperfine shift that we aim to measure. The second term arises from the spin-echo $\pi$-pulse not being effective, so that the hyperfine interaction with the electron spin cancels. The third term arises from the $\pi/2$-pulse not exciting the nuclear spin and the $\pi$-pulse creating some coherence, analogous to performing a Ramsey during the second half of the sequence (when the electron is in the $m_s = +1$ state). These spurious frequencies are easily identifiable in the signal, as $\omega_\mathrm{L}$ and $A^{+}_{j}$ are typically $>100$ kHz, while $\deltah{j}$ ($\sim 1 - 50$ kHz) is tightly bound by the bandwidth of the RF pulses ($\sim 1$ kHz). Furthermore, the second and third terms decay quickly ($\tstar$-limited), as the spin-echo is not effective. Hence, any signal remaining after $\sim 10$ ms contains only the $\deltah{j}$ term of interest.

\subsubsection{Electron-state-dependent dephasing}\label{seq:esd_dephasing}
\begin{figure}
	\begin{center}
		\includegraphics[width=0.9\columnwidth]{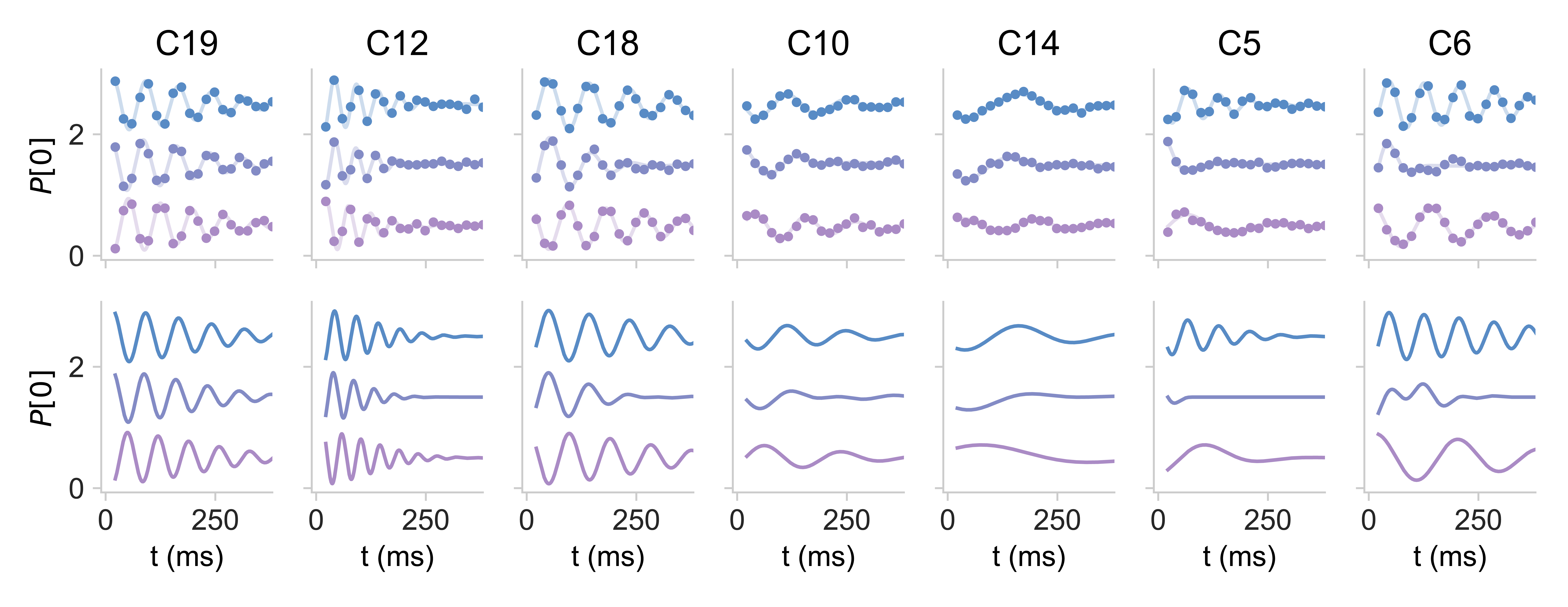}
	\end{center}
\caption{\textbf{Simulating electron-state-dependent dephasing for various nuclear spins} Top: Experimental data (as in Supplementary Fig. \ref{fig_ext:Electron_state_dependent}) for seven different nuclear spins after initialising the spin bath `up' (blue), `down' (pink) or in a `mixed' (purple) state (offset for clarity). The `mixed' signals exhibit a highly spin-dependent coherence decay, while polarising the bath leads to increased coherence and a shift in the frequency. The experimental data are corrected for difference in polarisation direction for selective and global initialisation (which leads to a minus sign). Bottom: Analytical model for the signal of the spins, obtained by evaluating Supplementary Eqs. \ref{eq:ESD_cosine_product} and \ref{eq:ESD_cosine_product_polarised}, reproducing both the observed coherence decay and frequency shifts qualitatively, taking into account the couplings to all other 49 spins. The amplitude, phase, frequency and (limiting) $\ttwo$-decay of the base signal are extracted from a fit to the `up' data.}
    \label{fig:Electron_state_dependent}
\end{figure}

At finite $\Bz$ field, a particular dephasing mechanism that we denote electron-state-dependent dephasing (ESD), prevents nuclear spins from attaining the full $\ttwo$ coherence time. The magnitude of this effect is highly spin-dependent and can be accurately modelled using the nuclear-spin interactions obtained here. As detailed in Ref. \cite{abobeih_atomicscale_2019}, the spin-spin couplings ($C_{ij} \neq C^{+}_{ij} \neq C^{-}_{ij}$), weakly depend on the electron spin state, resulting in an effective frequency shift of all spin-spin couplings during the electron-nuclear double resonance sequence (Supplementary Eq. \ref{eq:U_ENES}): 
\begin{equation}
    \Delta C_{ij} = \frac{1}{2}\left( C^{+}_{ij} - C^{-}_{ij} \right) \,.
\end{equation}
The dominant contribution is due to a change in the nuclear quantisation axes \cite{abobeih_atomicscale_2019}:
\begin{equation}\label{eq:quant_axes}
    \Delta C_{ij}  \approx \frac{(\Azx^{(i)}+ \Azx^{(j)}) C^{(ij)}_{zx} +  (A_{zy}^{(i)} + A_{zy}^{(j)}) C^{(ij)}_{zy}  }{\gc \Bz} \, ,
\end{equation}
with $A_{z \alpha}^{(i)}$ and  $C^{(ij)}_{z\alpha}$ the perpendicular components of the hyperfine and nuclear-nuclear dipole tensor, respectively. The key insight is that the couplings of spin $j$ to the network change between the first and second half of the spin-echo sequence. With the surrounding spin bath in a mostly mixed state (Supplementary Eq. \ref{eq:DNP_initial_state}), the quasi-static noise is not completely eliminated by the spin echo, as the $\Delta C_{ij}$-terms do not cancel. Typically these terms do not exceed $\sim 2$ Hz (see also Ref. \cite{abobeih_atomicscale_2019}), but for some spins with strong couplings to the electron spin ($A_{\perp} \gg 10$ kHz), the effect leads to a loss of coherence comparable to $\tstar$ decay (see Supplementary Fig. \ref{fig:Electron_state_dependent},  C5 and C6).

To model the effect on the measured signal, we consider two limiting cases: one where the surrounding spin bath is in a completely polarised state (denoted by `up' or `down' in Supplementary Fig. \ref{fig:Electron_state_dependent}) and one in which it is in a mixed state. Considering the effect of $N$ nuclear spins on a target nuclear spin $j$, we find that the signal frequency gets shifted when the bath is polarised ($p_i = \pm 1 \, \forall \, i$ in Supplementary Eq. \ref{eq:DNP_initial_state}). 
\begin{equation} \label{eq:ESD_cosine_product_polarised}
    \expval{\Iz{j}} =  \frac{1}{2} p_j \cos{\left[ 2\pi (\deltah{j} \pm \phi_j) \, t  \right]}
\, , \,\,\,\,\,
\phi_j = \sum_i^N \frac{\Delta C_{ij}}{2}\,,
\end{equation}
where the $\pm$-sign is given by the direction of the bath polarisation (`up' or `down'). 

Next, assuming a mixed state for the bath spins ($p_i = 0 \, \forall \, i$ in Supplementary Eq. \ref{eq:DNP_initial_state}), we find:
\begin{equation} \label{eq:ESD_cosine_product}
    \expval{\Iz{j}} =  \frac{1}{2} p_j \cos{\left( 2\pi \deltah{j} \, t  \right)} \,
    \prod_{i=1}^{N} \cos{\left( 2\pi \, \frac{\Delta C_{ij}}{2}  \, t  \right)} 
\end{equation}
In this case, the coupled spins will cause frequency beatings on the signal, leading to a decay:
\begin{equation} \label{eq:ESD_decay}
    \expval{\Iz{j}} \approx  \frac{1}{2} p_j \cos{\left( 2\pi \deltah{j} \, t  \right)} \,
    e^{-\left(t/T_{2,\mathrm{ESD}}\right)^2} \,
\end{equation}
with characteristic decay time:
\begin{equation}
    T_{2,\mathrm{ESD}} = \sqrt{  \frac{2}{\sum^N_{i=1} \left( 2\pi \, \frac{\Delta C_{ij}}{2} \right)^{2}}}
\end{equation}

Supplementary Fig. \ref{fig:Electron_state_dependent}b-e shows the experimental observation of the ESD effect for seven selected spins in both polarised and mixed spin bath conditions. These conditions are achieved by using either the global PulsePol sequence or a selective SWAP initialisation of the target nuclear spin (Methods). The obtained signals display significant coherence and frequency variation between spins and depend strongly on the state of the bath. 

To model this behaviour, we first extract the frequency, amplitude, phase and coherence time of the `up' data. Next, we calculate how the signal should change for different bath states due to the ESD effect, using Supplementary Eqs. \ref{eq:ESD_cosine_product} and \ref{eq:ESD_cosine_product_polarised} to generate a spin-specific model. To this end, the $\Delta C_{ij}$ for each of the spins are calculated according to Supplementary Eq. \ref{eq:quant_axes}, based solely on the known spin positions (assuming dipolar hyperfine coupling). Supplementary Fig. \ref{fig:Electron_state_dependent}f-i shows the modelled signal, for which we observe good qualitative agreement for each of the spins. The difference in initial phase between `up' and `mixed' data for some spins (Supplementary Fig. \ref{fig:Electron_state_dependent}b,e) is due to a difference in polarisation direction between the selective and global initialisation sequences. Furthermore, using experimentally determined values of $\Delta C_{ij}$ \cite{abobeih_atomicscale_2019} (since we know Supplementary Eq.\ref{eq:quant_axes} to be approximate), further diminishes the discrepancy between model and data.

Typically, spins that are more strongly coupled to the electron spin (e.g. panel b) show a quick decoherence behaviour for unpolarized environments(see Supplementary Eq. \ref{eq:quant_axes}). For more weakly coupled spins, the decoherence becomes determined by the basic $T_2$-echo time. For such more weakly coupled spins as well for highly polarised baths, we expect multiple refocussing pulses can further enhance coherence and therefore the resolution of the sequence.

\subsubsection{Electron-nuclear double resonance spectroscopy}
\begin{figure}
	\begin{center}
		\includegraphics[width=0.65\columnwidth]{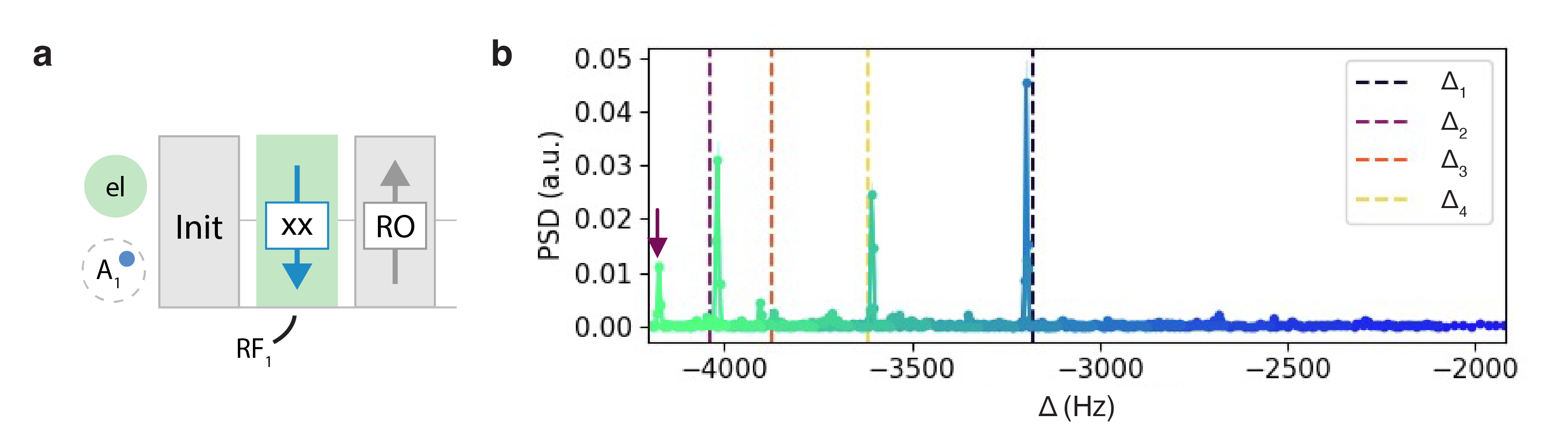}
	\end{center}
\caption{\textbf{Direct nuclear spectroscopy.} \textbf{a} Experimental sequence for performing double resonance spectroscopy directly with the electronic quantum sensor (as in Fig. 3). The pulses and readout can be tuned so that only signal from spins in a frequency region of interest ($A$) is picked up. This region of interest is then swept over a larger region to create a stitched spectrum. \textbf{b} Stitched spectrum (color shading shows individual datasets) as a function of hyperfine shift $\deltah{}$, demonstrating transform-limited spectral resolution ($\sim 10$ Hz). Dashed lines denote the estimated $\deltah{i}$ of four previously characterized spins in the frequency region. The purple arrow denotes a spurious alias, corresponding to $\deltah{1}$. The error is smaller than the data points (calculated according to Ref. \cite{hoyng_error_1976}).}
    \label{fig:ENES_spectroscopy}
\end{figure}

Next, we discuss how to use the electron-nuclear double resonance sequence to perform \emph{direct} high-resolution spectroscopy of nuclear spins. Since the sequence only enhances coherence for spins resonant with the RF-pulses, we take multiple data sets at varying RF frequencies and stitch them together to create a larger scan. To this end, we implement the sequence sketched in Supplementary Fig. \ref{fig:ENES_spectroscopy}a. For the spectroscopy data in Supplementary Fig. \ref{fig:ENES_spectroscopy}b, we sweep frequency $\mathrm{RF}_1 \approx \omega_{\mathrm{L}} - \deltah{}$ from $\sim 434 - 436$ kHz ($\mathrm{RF}_1$ in the `xx'-block), keeping the pulse Rabi frequency at $0.38$ kHz and updating the DD readout parameters as (`RO'-block in Supplementary Fig. \ref{fig:ENES_spectroscopy}):
\begin{equation}
    \tau_{\mathrm{DD}} = \frac{k}{4 \left(\omega_{\mathrm{L}} - \deltah{} / 2 \right)} \, ,
\end{equation}
with $k = 41$ the DD resonance order and keeping the number of pulses fixed to $N = 208$. The update rule ensures that the readout remains resonant with the frequency of interest ($A_1$) \cite{taminiau_detection_2012, degen_quantum_2017}. For each frequency, we sweep the double resonance evolution time $t$ (Supplementary Fig. \ref{fig:ENES_spectroscopy}a) up to $\SI{200}{\milli \second}$  (bandwidth of $200$ Hz) and compute the PSD, which we stitch together to create the large bandwidth, high resolution spectroscopy data (Supplementary Fig. \ref{fig:ENES_spectroscopy}b).

The spectral region interrogated in Supplementary Fig. \ref{fig:ENES_spectroscopy}b is known to contain four nuclear spins (dotted lines), three of them being visible in the signal with almost transform-limited linewidth ($\sim 10$ Hz). The amplitude of the signal is determined by a combination of polarisation efficiency (see Supplementary Eq. \ref{eq:ENES_signal}) and readout fidelity \cite{taminiau_detection_2012, abobeih_onesecond_2018}. The colored arrow denotes suspected aliasing, which can be easily mitigated by increasing the sampling bandwidth.

The double resonance spectroscopy presented here can be readily implemented for other color center platforms \cite{debroux_quantum_2021, sipahigil_integrated_2016, bourassa_entanglement_2020, durand_broad_2021} to interrogate the nuclear spin environment with high spectral resolution. By using DDrf readout sequences, the protocol is further simplified and can be used for nuclear spins with small perpendicular hyperfine coupling \cite{bradley_tenqubit_2019}.

\subsubsection{Electron-nuclear double resonance of the 50-spin-network}

\begin{figure}
	\begin{center}
		\includegraphics[width=1\columnwidth]{figure_other_ENES_sequences_v2-01.png}
	\end{center}
\caption{\textbf{Electron-nuclear double resonance of the spin-network}. \textbf{a-d} Pulse sequences used in this work to extract $\deltah{i}$ for all spins in Supplementary Table \ref{tab:spin_overview}. We try to use chains of minimum length to limit experimental time and complexity. The exact sequence that was used is denoted in the `Readout' column in Supplementary Table \ref{tab:spin_overview}.}
    \label{fig:all_ENES_data}
\end{figure}

We perform the electron-nuclear double resonance sequence on all known spins in the network. As we can only access a number of spins directly, we implement the electron-nuclear double resonance block with a spin chain of varying length (See Supplementary Fig. \ref{fig:all_ENES_data}). The used sequence for each spins can be found in Supplementary Table \ref{tab:spin_overview}.

To retrieve the absolute signal frequency from undersampled data, we take at least two data sets with different bandwidths and sampling rates \cite{boss_quantum_2017}. Next, we correct for aliasing by minimizing the mean squared error between the multiple measurements, and selecting the most likely alias. We use prior knowledge of the $m_s = -1$ frequency of the spin to limit this analysis to a frequency range of 400 Hz around the expected resonance.

The values obtained for $\deltah{i}$ are shown in Supplementary Table \ref{tab:spin_overview}, where the error denotes the weighted error on the mean of multiple measurements (whose error is determined from an exponentially decaying fit). For many of the previously characterised spins \cite{abobeih_atomicscale_2019}, we find good agreement with Ramsey measurements. However, for some spins (most notably C19), we find a deviation that cannot be explained by off-axis fields or a correction due to the (reported) Fermi contact term \cite{nizovtsev_nonflipping_2018}. Further research is needed to identify the discrepancy between the Ramsey method used in Ref. \cite{abobeih_atomicscale_2019} and values reported in this work.

\clearpage
\subsection*{Supplementary Note 5: Network reconstruction algorithm} \label{sec:sup_algorithm}
\setcounter{subsubsection}{0}

The network mapping procedure outlined in the Methods section relies on the (pseudo) function CheckVertex($w,T$). Here, we present pseudo-code for the procedure of checking whether vertex $w$ has already been mapped in $T$.
\newline
\begin{algorithmic}
\Function{CheckVertex}{w, T} \Comment{Checks if $w$ was already characterised and returns the duplicate vertex}
    \newline
    \State unique = \texttt{True} \Comment{Boolean, keeps track whether vertex $w$ is unique}
    \State duplicate = \texttt{None} \Comment{Specifies duplicate vertex if applicable, otherwise \texttt{None}}
    \newline
    \For{k in T}
        % Check by frequency A
        \If{$|A_w - A_k| < \sigma_A$} \Comment{Compare frequency of $w$ to mapped vertices, $\sigma_A$ denotes uncertainty}
            \State unique = \texttt{False} \Comment{$w$ and $k$ might be duplicates}
            \newline
            
            \State $V_0^w  = \{w\}$ \Comment{Create a spanning tree $T_w$ with root $w$}
            \State $i = 0$ \Comment{Keeps track of depth of search}
            \State $j = 0$ \Comment{Keeps track of number of equivalent edges}
            \While{\textbf{not} unique \textbf{and} duplicate == \texttt{None}  \textbf{and} $i < $ maxdepth} %\Comment{Compare for a maximum depth starting from $w$}                
                \For{each vertex $v^w \in V_i^w$ and $v^k \in V_i^k$} \Comment{Find equivalent vertex in spanning tree $T_k$ with root $k$}
                    \State $\deltah{}$ = MeasureDelta($v^w$)  \Comment{Electron-nuclear double resonance measurement}
                    \If{$|\deltah{} - \deltah{k}| > \sigma_\Delta$} \Comment{$\sigma_\Delta$ denotes uncertainty}
                        \State unique = \texttt{True} \Comment{$w$ is unique, different $\deltah{}$}
                    \EndIf
                    % Confirming
                    \If{$|\deltah{} - \deltah{k}| < \sigma_\Delta$ \textbf{and} $\sigma_\Delta <$ threshold } 
                        \State unique = \texttt{False} \Comment{$w$ is equal to $k$, similar $\deltah{}$ unlikely}
                        \State duplicate = $k$
                    \EndIf

                    \For{each vertex $r^k$ $\in V_{i+1}^k$} \Comment{Get known neighbours of $k$ from $T_k$}
                        \State C = MeasureCoupling($v^w,A_r^k$) \Comment{Between vertex $v^w$ in $T_w$ and the frequency of $r^k$ in $T_k$}
                        \If{$|C - C_{vr}^k| > \sigma_C$} \Comment{Compare to edge in $T_k$}
                            \State unique = \texttt{True} \Comment{$w$ is unique, different spanning tree}
                        \EndIf
                        \If{$|C - C_{vr}^k| < \sigma_C$ \textbf{and} $\sigma_C <$ threshold} \Comment{Measurement has reasonable uncertainty}
                            \State $j = j + 1$  \Comment{Another edge of the spanning tree coincides}
                            \State create $r^w$ in $T_w$ \Comment{Expand $T_w$}
                            \State $A_r^w = A_r^k$
                            \State $C_{vr}^w = C$ 
                            \State add $r^w$ to $V_{i+1}^w$ in $T_w$
                        \EndIf
                    \EndFor
                \EndFor
                \State $i = i + 1$
            \EndWhile
            \If{$j > $ equaledges }
                \State unique = \texttt{False} 
                \State duplicate = $k$ \Comment{$w$ and $k$ are the same vertex, similar spanning tree unlikely}      
            \EndIf
        \EndIf
    \EndFor
    \newline
    \State \Return unique, duplicate
\EndFunction
\newline
\end{algorithmic}

The function starts by comparing the ($\tstar$-limited) frequency $A_w$ of vertex $w$ to the frequencies of all mapped vertices in $T$. If the frequencies coincide within the measurement uncertainty ($\sigma_A \sim 100$ Hz) for a vertex $k$, we initiate a procedure to check whether $w$ is that vertex. We do this by comparing information we have on vertex $k$ and its surroundings to specific measurements taken from $w$. In particular, we measure the spanning tree $T_w$ with root $w$, and compare it to the spanning tree $T_k$ with root $k$ (for some maximum depth). If all couplings ($C_{ij}$, up to some threshold `equaledges') and hyperfine frequency shifts ($\deltah{i}$) of the two spanning trees are the same, we conclude $w$ and $k$ are the same vertex. If we measure a single deviation, we conclude $w$ must be unique. If the procedure is inconclusive, for example because the uncertainty of all measurements is large, `duplicate' remains \texttt{None}.

% \clearpage
% \subsection*{Supplementary References}
\onecolumngrid

% \bibliographystyle{naturemag}
% \bibliography{dissertation}